\documentclass[preprint,onecolumn,aps,nofootinbib]{revtex4}
\usepackage[T1]{fontenc}
\usepackage{graphicx}
\usepackage{amsmath}
\usepackage{amsfonts}
\usepackage{amssymb}
\usepackage[colorlinks=true,linkcolor=blue,urlcolor=blue,citecolor=red]{hyperref}
\usepackage[normalem]{ulem}

\begin{document}

\title{\boldmath Optical conductivity of holographic striped superconductors}
\author{Fang-jing Cheng $^{1,2}$}
\email{chengfj@ihep.ac.cn}
\author{Kai Li $^{2,3}$}
\email{lik@ihep.ac.cn}
\author{Yi Ling $^{2,3}$}
\email{lingy@ihep.ac.cn}
\author{Meng-He Wu$^{4}$}
\email{mhwu@njtc.edu.cn}

\affiliation{
    $^1$ School of Physics and Astronomy, Beijing Normal University, Beijing, 100875, China\\
    $^2$ 
    Institute of High Energy Physics, Chinese Academy of Sciences, Beijing 100049, China;\\
    $^3$ School of Physics, University of Chinese Academy of Sciences, Beijing 100049, China;\\
    $^4$ College of Physics and Electronic Information Engineering, Neijiang Normal University, Neijiang 641112, China.
}

\begin{abstract}
Based on a holographic model exhibiting Mott-like insulating behavior on an ionic lattice, we construct a holographic striped superconductor (SSC), which is characterized by the interplay between the charge density wave (CDW) and superconductivity (SC). We present the phase diagram on the $e$-$T$ plane, where the SSC phase emerges within an intermediate window of $e$, revealing both cooperative and competitive relations between the CDW order and SC order. For small charge $e$, the system undergoes the phase transition from metallic phase to CDW phase, and then to SSC phase upon cooling; while for large $e$, it transitions from metallic phase to SC phase directly, and then to SSC phase. We compute the optical conductivity of the system for these two cases and demonstrate its distinct behavior. It turns out that the imaginary part of the conductivity exhibits a divergent behavior below the critical temperature $T_{\mathrm{SSC}}$ along both pathways, so that the superconductivity of SSC phase is justified for the first time in the holographic framework. Along the CDW-to-SSC pathway the pseudogap and the pinning effect are inherited from the CDW phase, whereas along the SC-to-SSC pathway the spontaneous stripe strongly suppresses the coherent normal-fluid channel while the superfluid pole is enhanced. In addition, we demonstrate the commensurate lock-in behavior of the striped superconductor in the $p$-$T$ diagram. 
\end{abstract}

\maketitle
\tableofcontents

\section{Introduction}\label{sec:intro}
High-temperature superconductivity has long been a central subject in condensed matter physics~\cite{Bednorz:1986tc,Keimer:2015zqv}. Experimentally, many strongly correlated superconductors, with the cuprates---doped Mott insulators---as the most prominent examples, exhibit not only superconducting pairing but also spatially modulated modes dubbed pair density waves (PDW)~\cite{Hamidian2016PDW}. Superconducting phases with such stripe-like modulations are commonly referred to as striped superconductors (SSC)~\cite{Berg_2009,Agterberg_2020}. Physically, multiple orders in the SSC phase act on the same electrons near the Fermi surface in different ways: charge density wave (CDW) order reconstructs the charge distribution and can partially gap the Fermi surface~\cite{Gruner1988}; superconducting (SC) order organizes charge carriers into a condensate with a singular zero-frequency response; and the PDW order corresponds to superconducting pairing at a finite wave-vector. Their coupling, through which the orders compete, coexist, or reinforce one another, determines how the low-energy spectral weight is redistributed~\cite{Fradkin2015Intertwined,Basov2005Electrodynamics}. A complete theoretical description of this phenomenology is still absent because it must treat strong correlations, spatial modulation, and nonlinear couplings among several order parameters at the same time, beyond the reach of a simple weak-coupling or Landau-Fermi-liquid picture.

Gauge/gravity duality or the holographic gravity~\cite{Maldacena:1997re,Gubser:1998bc,Witten:1998qj} provides a powerful non-perturbative framework for such strongly coupled and spatially modulated systems, especially when a quasiparticle description is not available. In the context of holographic gravity, superconductivity is realized by the condensation of a charged scalar field in bulk which signals the spontaneous breaking of the $U(1)$ symmetry~\cite{Gubser:2008px,Hartnoll:2008vx,Hartnoll:2008kx}. To distinguish the resulting superfluid pole from the zero-frequency response caused by exact momentum conservation, a broad range of holographic lattices and momentum-relaxing mechanisms has been developed, yielding finite DC transport and Drude-like optical response in the metallic regime~\cite{Horowitz:2012ky,Horowitz:2012gs,Vegh:2013sk,Liu:2012tr,Donos:2012js,Horowitz:2013jaa,Ling:2013nxa,Blake:2013owa,Andrade:2013gsa,Donos:2013eha,Donos:2014uba,Donos:2014yya,Ling:2014laa,Ling:2014bda,Baggioli:2014roa,Ling:2016lis,Ling:2017naw,Wu:2018zdc,Baggioli:2021xuv,Wang:2021jfu}. In parallel, holographic CDW phases have been generated through the spontaneous breaking of translations which leads to the formation of spatially modulated charge densities, with extensive work on their collective modes, pseudogap-like suppression, pinning effect, and transport properties~\cite{Ooguri:2010xs,Donos:2011bh,Donos:2013wia,Donos:2013gda,Withers:2013loa,Withers:2013kva,Ling:2014saa,Andrade:2017cnc,Alberte:2017oqx,Amoretti:2017axe,Li:2018vrz,Ammon:2019wci,Song:2019rnf,Andrade:2020hpu,Ling:2020qdd,Amoretti:2021fch,Baggioli:2022pyb,Baggioli:2019abx,Baggioli:2020edn}. In the literature, two main mechanisms are available to generate such a CDW in holography. One is to couple a neutral pseudo-scalar $\chi$ to the Maxwell field through a topological term $\vartheta(\chi)F\wedge F$~\cite{Donos:2011bh,Donos:2013wia,Andrade:2017ghg}; the CDW is then necessarily accompanied by a current-density wave and the breaking of parity and time reversal symmetry. These are features that a real CDW need not have~\cite{Donos:2013gda}. The other one, as adopted in this paper, is to introduce two $U(1)$ gauge fields coupled through a dilaton, such that the CDW forms without any current-density wave, preserving both symmetries~\cite{Donos:2013gda,Ling:2014saa}.

At an early stage some holographic models on striped superconductivity and PDW order were constructed through the St\"uckelberg mechanism~\cite{Cremonini:2016rbd,Cremonini:2017usb,Cai:2017qdz}, where a single order parameter was applied to describe both charge modulation and superconducting condensation. Since the CDW order is rooted in the spontaneous breaking of the translational symmetry, while the SC order results from the $U(1)$ gauge symmetry breaking, in principle these orders may form separately and thus should be described by distinct order parameters. Based on these considerations new holographic models of striped superconductors were constructed in~\cite{Ling:2019gjy,Ling:2020qdd,Li:2024ybq}, where distinct order parameters are specified for CDW order and SC order individually such that the PDW order appears explicitly through the interplay between the CDW order and SC order. Furthermore, recently the holographic CDW models have been developed to reproduce the commensurate lock-in effect by incorporating the inhomogeneous lattices into the background~\cite{Andrade:2017leb,Krikun:2017cyw,Ling:2023ncu}---It is worth stressing that homogeneous lattices cannot capture the lock-in of the wave-vector in the context of holography~\cite{Andrade:2015iyf,Andrade:2018gqk,Ling:2020mwm}. In the lock-in state, the DC resistivity rises algebraically on cooling~\cite{Ling:2023ncu}, exhibiting behavior similar to that of the holographic Mott insulator constructed in Ref.~\cite{Andrade:2017ghg}. This locked CDW phase thus provides a Mott-like parent state for a holographic striped superconductor, analogous to the structure of actual high-temperature superconducting materials. The framework presented in \cite{Ling:2020mwm,Ling:2023ncu,Li:2024ybq} combines all these ingredients into a single model, namely building a striped superconductor with multiple order parameters over a Mott-like insulator, and thus becomes the appropriate setup close to actual high-temperature superconducting materials. Based on this framework, we intend to explore the transport behavior of SSC in this paper.

Optical conductivity provides an experimentally accessible probe of the interplay between charge order and superconductivity. However, the optical conductivity of holographic SSC in this setting has not been investigated in the literature due to the complicated numerical simulations involved\footnote{In Ref.~\cite{Ling:2019gjy}, the optical conductivity is demonstrated for a superconducting phase purely induced by a pre-existing CDW. It should be pointed out that the model does not contain the ordinary free charge such that it is not the ordinary SSC model which should have free charges to form PDW. In addition, an explicit ionic lattice is absent in that model. Thus the optical response of an SSC with finite charge carriers, lattice-induced momentum relaxation, and commensurate lock-in has not yet been systematically studied.}. Therefore, one of the main purposes of this paper is to justify the superconductivity of the holographic SSC models by computing the optical conductivity explicitly, and identifying the zero-frequency superfluid contribution after the phase transition. Furthermore, the regular finite-frequency spectrum also allows us to examine how its transport behavior is influenced by momentum relaxation, CDW pinning, and superconducting condensation. Investigating these optical signatures provides a way to connect holographic predictions with experimental observations in real materials. Thus, another goal of this paper is to gain insight into how charge order and lattice commensurability affect the transport properties in striped superconductors.

Based on the above motivations, we investigate the optical conductivity of holographic SSC in a model with an inhomogeneous ionic lattice in this paper. We examine the superconducting response and regular transport near the transitions into the SSC phase along both the CDW-to-SSC and SC-to-SSC pathways, comparing the unlocked and locked regimes. We will explicitly demonstrate that the imaginary part of the conductivity becomes divergent when transforming from CDW phase to SSC phase such that its superconductivity is justified. Specifically, this model contains independent CDW and SC order parameters, so the SSC phase, together with its PDW order, can be reached through two distinct pathways upon cooling: for small charge $e$ which implies weaker superconducting pairing, the CDW forms first and the superconductivity appears on a pre-existing charge-ordered background, while for larger $e$ the SC phase appears first and the CDW develops on a pre-existing superconducting background. Interestingly enough, we find that the transport behavior of SSC, particularly its zero-frequency behavior, strongly depends on the pathway it undergoes upon cooling. Along the CDW-to-SSC pathway, the transition is preceded by a strong suppression of the DC conductivity and by the formation of a pseudogap; the superconductivity then mainly adds a zero-frequency superfluid component on top of this already formed CDW optical spectrum, with only mild changes in the regular finite-frequency conductivity. Along the SC-to-SSC pathway, by contrast, the CDW order formed simultaneously with the transition strongly suppresses and broadens the residual near-DC normal response, while the superfluid pole remains present and is enhanced. In the locked regime, the regular low-frequency response is already controlled by explicit lattice scattering, and the additional spontaneous stripe leaves a much weaker finite-frequency signature.

The paper is organized as follows. Section~\ref{sec:setup} introduces the holographic setup with multiple order parameters, where the CDW and SC orders are implemented over the ionic lattice background separately. Section~\ref{sec:phase} demonstrates the phase diagrams of the system: the $e$-$T$ diagram displays two distinct pathways into the SSC phase upon cooling, while the $p$-$T$ diagram describes the wave-vector selection and commensurate lock-in effect of the SSC order as the amplitude of the lattice changes. Section~\ref{sec:sigma} focuses on the linear-response computation of the optical conductivity. We investigate it as a function of frequency near the phase transition and demonstrate it as a function of temperature in the near-DC limit. The distinct behavior of the conductivity is also demonstrated for two pathways in both unlocked and locked regimes. Section~\ref{sec:conclusion} summarizes the main results and discusses possible extensions. The equations of motion are collected in Appendix~\ref{app:eoms}, and the temperature dependence of the order parameters and charge densities in Appendix~\ref{app:profiles}.

\section{Holographic Setup}\label{sec:setup}
We consider a four-dimensional holographic model containing the metric $g_{\mu\nu}$, a dilaton field $\Phi$, two $U(1)$ gauge fields $A_\mu$ and $B_\mu$, and a complex scalar field $\Psi$ charged under $A_\mu$.
The action is
\begin{equation}
\begin{aligned}
S = & \frac{1}{2\kappa^2} \int d^4x \sqrt{-g} \bigg[ R - \frac{1}{2}(\nabla\Phi)^2 - V(\Phi) - \frac{1}{4} Z_A(\Phi)F^2 \\
    &- \frac{1}{4} Z_B(\Phi)G^2 - \frac{1}{2} Z_{AB}(\Phi)FG - \left|\left(\nabla - i e A\right)\Psi\right|^2 - m_v^2\,\Psi\Psi^* \bigg],
    \label{eq:action}
\end{aligned}
\end{equation}
where the potential and coupling functions are
\begin{equation}
V(\Phi) = -\frac{1}{L^2} + \frac{1}{2} m_s^2 \Phi^2, \quad Z_A(\Phi) = 1 - \frac{\beta}{2} L^2 \Phi^2, \quad Z_B(\Phi) = 1, \quad Z_{AB}(\Phi) = \frac{\gamma}{\sqrt{2}} L \Phi.
\end{equation}
Here $R$ is the Ricci scalar, $g$ the metric determinant, $\kappa$ the gravitational coupling, $l=\sqrt{6}L$ the AdS$_4$ radius, and $m_s$, $m_v$ the masses of $\Phi$ and $\Psi$, respectively. The field strengths of the gauge fields are $F=dA$ and $G=dB$. In this model, a nonzero $\Phi$ breaks translational symmetry spontaneously and is the order parameter of the charge density wave (CDW) phase, while a nonzero $\Psi$ breaks $U(1)$ symmetry spontaneously and is the order parameter of the superconducting (SC) phase. Usually, we may write this complex scalar field as $\Psi = \eta e^{i\theta}$. The phase $\theta$ is set to zero by gauge fixing for the background solutions such that the magnitude $\eta=|\Psi|\neq0$ is the signature of the SC condensation. In Refs.~\cite{Ling:2014saa, Ling:2019gjy}, the gauge field $B_\mu$ is treated as the electromagnetic field. But by this setup $B_\mu$ is always zero until the translational symmetry is broken spontaneously. Even after the symmetry is broken, the free charge $\rho_{B}^{(0)}$ is always zero and only the CDW is present in the background. In this paper, by contrast, we insist that both free charges and the CDW should be present, as in real materials. Thus we treat $A_\mu$ as the physical electromagnetic field dual to the boundary electric current; the secondary gauge field $B_\mu$ is kept source-free and serves as a responsive channel for the spontaneous charge modulation.

Since this model has been considered many times in the holographic literature, for conciseness we list the equations of motion in Appendix~\ref{app:eoms}. To construct backgrounds with an ionic lattice along the $x$ direction, we adopt the following ansatz
\begin{equation}
\begin{aligned}
ds^2 &= \frac{1}{z^2}\left[-(1-z)p(z)Q_{tt}dt^2 + \frac{Q_{zz}}{(1-z)p(z)}dz^2 + Q_{xx}(dx+z^2 Q_{xz}dz)^2 + Q_{yy}dy^2\right],\\
A_t &= (1-z)\mu(x)\,a,\quad B_t=(1-z)\,b,\quad \Phi=z\,\phi,\quad \eta=z\,\zeta,
\end{aligned}
\end{equation}
where $p(z) = 4\left(1+z+z^2-\frac{\mu^2 z^3}{16}\right)$ and $Q_{tt},Q_{zz},Q_{xx},Q_{yy},Q_{xz},a,b,\phi,\zeta$ are functions of $x$ and $z$. For numerical convenience the horizon is set at $z=1$ and the AdS boundary at $z=0$. When $Q_{tt}=Q_{zz}=Q_{xx}=Q_{yy}=a=1$ and $Q_{xz}=b=\phi=\zeta=0$, the background reduces to the AdS-RN black brane in the absence of the ionic lattice. Throughout this paper, we set $l^2=6L^2=1/4$, $m_s^2=m_v^2=-2/l^2=-8$, $\beta=-450$, and $\gamma=37.3$, working in the large-$N$ limit $l^2/2\kappa^2\gg1$. We take $\mu$ as the scaling unit, in terms of which the Hawking temperature is
\begin{equation}
\frac{T}{\mu} = \frac{48-\mu^2}{16\pi\mu},
\end{equation}
which remains valid even in the presence of the ionic lattice. The explicit ionic lattice is introduced through a spatially modulated boundary chemical potential for $A_\mu$,
\begin{equation}
\mu(x) = \mu\,(1+\lambda\cos kx),
\end{equation}
where $\lambda$ and $k$ are the amplitude and wave-vector of the lattice, respectively. We solve the equations of motion numerically, treating the Einstein equations with the DeTurck method~\cite{Figueras_2011,Adam_2012}. Throughout this paper we fix $k/\mu=2$, and consider the scalar charge $e$ and the lattice amplitude $\lambda$ as two adjustable parameters. Near the AdS boundary, the electromagnetic field behaves as $A_t \simeq \mu(x) - \rho_A(x)\,z + \cdots$, where $\rho_A(x)$ is identified as the charge density of the dual current. It can be further expanded as $\rho_A(x) = \rho_A^{(0)} + \rho_A^{(1)}\cos(kx) + \cdots$, with the uniform component $\rho_A^{(0)}$ measuring the free carriers; the temperature dependence of these components over different backgrounds is collected in Appendix~\ref{app:profiles}.

In addition, this work focuses on the striped superconducting (SSC) phase, where both the CDW order $\Phi$ and the SC order $\eta$ are nonzero. Of course, since these two orders are characterized by different order parameters, they may exist independently, leading to the sole CDW phase or SC phase, implemented by setting the other order parameter to zero. The above cases have already been studied in detail~\cite{Ling:2023ncu, Li:2024ybq}, but to keep the paper self-contained, we briefly review the main results in the next two subsections.

\subsection{The CDW phase with ionic lattice}\label{subsec:setup_cdw}
The CDW phase in the absence of the SC order corresponds to $\Phi \ne 0$ but $ \eta=0$. Over an ionic lattice background, the coupling $Z_A(\Phi)$ gives the dilaton an effective mass and may induce the instability of the background once this mass violates the Breitenlohner--Freedman bound. To estimate the critical temperature for this instability, one may consider the linear perturbations over the background by turning on the dilaton field and the gauge field $B_\mu$ as
\begin{equation}
\delta\Phi = \delta\Phi^{\tilde p}(x,z)\,e^{i\tilde p x}, \qquad \delta B = \delta B^{\tilde p}(x,z)\,e^{i\tilde p x},
\end{equation}
where $\tilde p$ is the CDW wave-vector and ranges from $0$ to $k$. Solving the resulting generalized eigenvalue problem gives rise to the critical temperature $T_c(\tilde p)$ of the CDW for each wave-vector $\tilde p$~\cite{Ling:2023ncu}.

The critical temperature $T_c$ for the phase transition is identified as the highest temperature at which the phase transition occurs, since the CDW forms most readily there. The corresponding wave-vector is denoted as $\tilde p_{\max}$. In the presence of the ionic lattice with wave-vector $k$, it is interesting to notice that the lock-in effect may be observed, as revealed in~\cite{Andrade:2017leb,Ling:2023ncu}. This effect indicates that once the lattice amplitude $\lambda$ is large enough, the ratio of $\tilde p_{\max}$ to the lattice wave-vector $k$ approaches a rational value. Such pinning of a spontaneous modulation to a rational multiple of the host lattice is known as commensurate lock-in, and it has long been studied for density waves in real materials~\cite{Gruner1988,McMillan1976,Bak1982}. For the parameters adopted in this paper, we find it eventually locks at $\tilde p_{\max}=k/2$, which then becomes the most stable CDW mode.

Once the critical temperature $T_c$ for an unstable mode is estimated by the perturbative analysis, we then solve the full nonlinear equations to obtain the CDW background numerically. Near the boundary $z=0$, the dilaton field asymptotically behaves as $\Phi \simeq \phi_1 z + \phi_2 z^2 + \cdots$. According to the holographic dictionary, $\phi_1$ is understood as the source and $\phi_2$ as the response. We are interested in spontaneous breaking of the translational symmetry, so the source is set to zero, $\phi_1=0$, and the subleading coefficient $\phi_2(x)$ becomes the CDW order parameter. Throughout this paper we will consider the solutions with the commensurate wave-vector $\tilde p=k/2$, where the period of the CDW is twice that of the lattice and the CDW eventually locks once the lattice amplitude is large enough. Then numerically we find the order parameter expands over the half-integer harmonics of $k$,
\begin{equation}
\phi_2(x) = \phi_2^{(1/2)} \cos\!\left(\frac{k}{2}x\right) + \phi_2^{(3/2)} \cos\!\left(\frac{3k}{2}x\right) + \cdots .
\label{phi2}\end{equation}
The temperature dependence of $\phi_2^{(1/2)}$ is collected in Appendix~\ref{app:profiles}.

\subsection{The SC phase with ionic lattice}\label{subsec:setup_sc}
In the pure SC state we turn off the charge density wave by setting $\Phi=0$, and look for nonzero solutions for the charged scalar $\eta$. Since $\Phi=0$ makes the coupling $Z_{AB}(\Phi)$ vanish, the secondary gauge field $B_\mu$ decouples and vanishes identically, leaving a single gauge field $A_\mu$ coupled to the charged scalar over the ionic-lattice background.

The superconductivity forms in the standard way: below some critical temperature the charged scalar $\eta$ becomes unstable and condenses, breaking the $U(1)$ symmetry of $A_\mu$~\cite{Gubser:2008px,Hartnoll:2008vx}. At the linear level the analysis parallels that of the CDW: we turn on a small perturbation of $\eta$ on the ionic-lattice background, $\delta\eta=\delta\eta^{p}(x,z)\,e^{ipx}$, insert it into the equations of motion, and recast the result as a generalized eigenvalue problem whose solution gives the superconducting critical temperature~\cite{Li:2024ybq}.

We then solve the full nonlinear equations with backreactions to obtain the background with SC condensation. Near the boundary the scalar behaves as $\eta \simeq \eta_1 z + \eta_2 z^2 + \cdots$, where $\eta_1$ is the source and $\eta_2$ the response. We set $\eta_1=0$ so that the $U(1)$ symmetry is broken spontaneously, and $\eta_2$ is treated as the SC order parameter. The lattice on $A_\mu$ modulates the condensate, so $\eta_2$ expands over the integer harmonics of $k$,
\begin{equation}
\eta_2(x) = \eta_2^{(0)} + \eta_2^{(1)} \cos(kx) + \eta_2^{(2)} \cos(2kx) + \cdots.
\label{eta2}\end{equation}
Here $\eta_2^{(0)}$ is the zero-momentum component of the condensate, while $\eta_2^{(1)}$ is the finite-wave-vector component induced directly by the ionic lattice. We stress that in the pure SC phase these finite-wave-vector components are merely imprinted by the explicit lattice and do not constitute a PDW order, which requires a spontaneous modulation of the condensate; such a spontaneous component emerges only in the SSC phase discussed below. Similarly, the temperature dependence of these components is collected in Appendix~\ref{app:profiles}.

\section{Phase Diagrams and Commensurate Lock-in}\label{sec:phase}
The striped superconducting (SSC) phase is formed when both the CDW order $\Phi$ and the superconducting order $\eta$ are turned on. Our goal is to understand how these two orders compete and cooperate during the course of coexistence. For this purpose we consider their temperature behavior with the change of two parameters $e$ and $\lambda$. The first is the scalar charge $e$, which sets how easily the superconductor condenses: the SC critical temperature increases with $e$, while the CDW does not couple to $e$, so varying $e$ shifts the balance between these two orders. The second is the lattice amplitude $\lambda$, which controls the commensurate lock-in: the $p$-$T$ phase diagram displays how, once $\lambda$ is large enough, the striped order developing on the superconducting background locks onto the lattice, in contrast to the pure CDW lock-in studied in Ref.~\cite{Ling:2023ncu}.

\subsection{The $e$-$T$ phase diagram}\label{subsec:eT}
\begin{figure}[htp]
    \centering
    \begin{minipage}{0.49\textwidth}\centering
        \includegraphics[width=\textwidth]{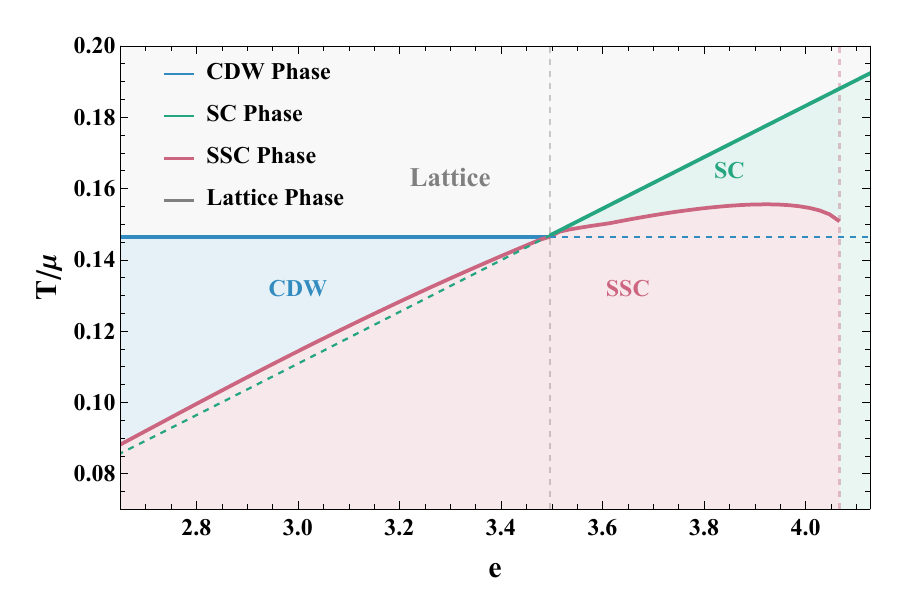}\end{minipage}\hfill
    \begin{minipage}{0.49\textwidth}\centering
        \includegraphics[width=\textwidth]{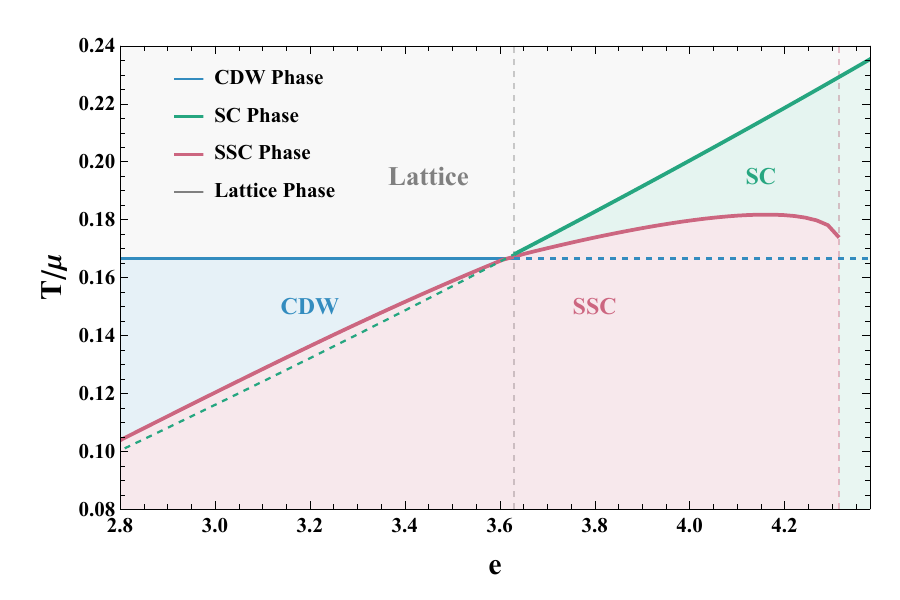}\end{minipage}

    \vspace{0.05cm}

    \begin{minipage}{0.49\textwidth}\centering
        \includegraphics[width=\textwidth]{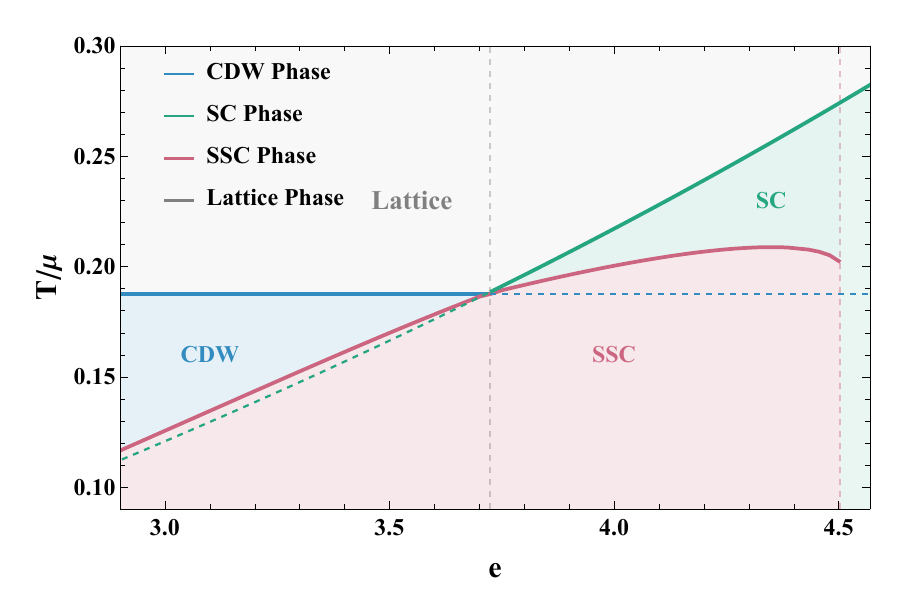}\end{minipage}\hfill
    \begin{minipage}{0.49\textwidth}\centering
        \includegraphics[width=\textwidth]{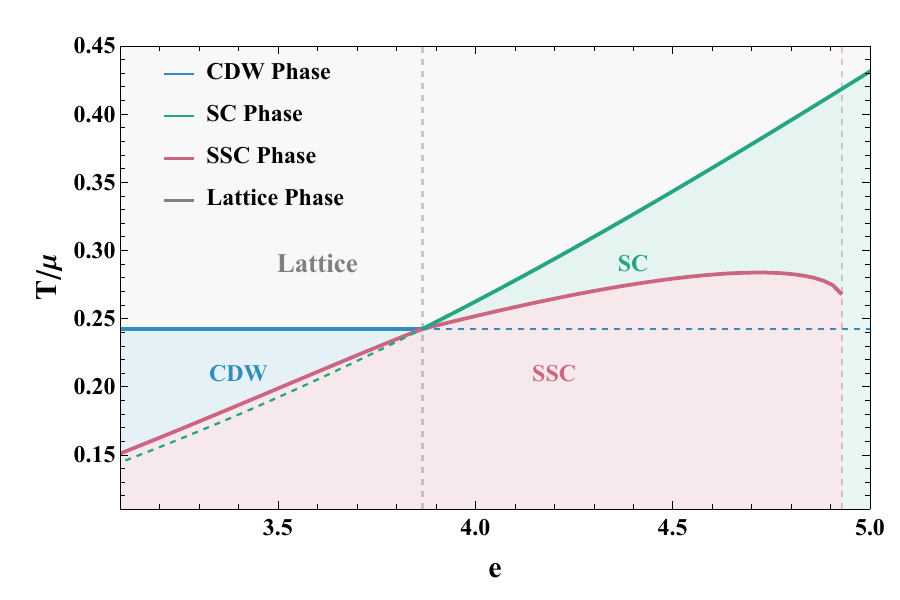}\end{minipage}

    \caption{Phase diagrams in the $e$-$T$ plane for different lattice amplitudes: $\lambda=0.4$ (top left), $\lambda=1$ (top right), $\lambda=1.3$ (bottom left) and $\lambda=1.8$ (bottom right). The vertical dashed lines in gray mark the position of $e_\ast$ at which $T_{\mathrm{CDW}}=T_{\mathrm{SC}}$. The vertical dashed lines in pink mark the value of $e$ where the SSC phase terminates. The remaining dashed lines indicate metastable states.}
    \label{fig:four_phase_diagrams}
\end{figure}

We plot the critical temperatures of three phases, namely the CDW, SC and SSC phases, as a function of the charge $e$ with a fixed lattice amplitude $\lambda$ in Fig.~\ref{fig:four_phase_diagrams}. Since the pure CDW forms with $\eta=0$, the charge $e$ does not enter its equations of motion, so $T_{\mathrm{CDW}}$ is independent of $e$ and appears as a horizontal line on the $e$-$T$ plane; on the other hand, the critical temperature of the SC $T_{\mathrm{SC}}$ continuously increases with the charge $e$, since a larger charge makes the scalar condense more readily. These two lines cross at $e_\ast$, where $T_{\mathrm{CDW}}=T_{\mathrm{SC}}$. With decreasing temperature, for $e<e_\ast$ the translational symmetry is spontaneously broken first, so that the CDW forms, while for $e>e_\ast$ the $U(1)$ gauge symmetry is broken first, so that the SC condenses.

For a given charge $e$, the SSC develops in two steps when the system is cooled down from the metallic phase with the lattice. Firstly, the system undergoes a phase transition to form the order with the higher critical temperature---the CDW for $e<e_\ast$ and the SC for $e>e_\ast$. Secondly, on the CDW or SC background, we lower the temperature further and apply the same linear-instability analysis to find the critical temperature $T_{\mathrm{SSC}}$ at which the SSC order starts to develop. The result is illustrated in Fig.~\ref{fig:four_phase_diagrams}. In this figure, it is remarkable to observe that for $e>e_\ast$, $T_{\mathrm{SSC}}$ always lies below $T_{\mathrm{SC}}$ but above $T_{\mathrm{CDW}}$. For instance, when $e=3.9$, the SC condenses first and then the SSC develops over the superconducting background at $T_{\mathrm{SSC}}>T_{\mathrm{CDW}}$. The situation for $e<e_\ast$ is similar: $T_{\mathrm{SSC}}$ always lies below $T_{\mathrm{CDW}}$ but above $T_{\mathrm{SC}}$. In both cases, forming the second order on the background of the first is easier than forming it over the bare lattice. This reflects the cooperation between the CDW and SC orders, which may be understood as follows. Both instabilities are driven by the gauge field, through the $e^2A^2\eta$ term for the superconducting scalar and the $Z_A'(\Phi)F^2$ term for the dilaton. We observe that whichever order forms first raises the charge density $\rho_A$ (Appendix~\ref{app:profiles}), and the stronger field lowers the effective mass of the second order, thus lifting its onset above the bare value.

Next we present another important fact that is observed in Fig.~\ref{fig:four_phase_diagrams}. As the charge $e$ becomes larger and approaches some value marked by the vertical dashed line in pink in the figure, no solution can be found for the SSC anymore. It means the translational symmetry cannot be broken spontaneously once the superconducting condensate becomes strong enough. This indicates a competitive relation between the SC and CDW orders: at large $e$ the SC grows strong enough to suppress the formation of the CDW. In addition, with the increase of the lattice amplitude, we find that all the critical temperatures rise, and that the value of $e$ which prevents the SSC order from forming becomes larger, as illustrated in Fig.~\ref{fig:four_phase_diagrams}.

\subsection{The $p$-$T$ phase diagram and commensurate lock-in}\label{subsec:pT}
\begin{figure}[htp]
\centering
\begin{minipage}{0.32\textwidth}\centering
\includegraphics[width=\textwidth]{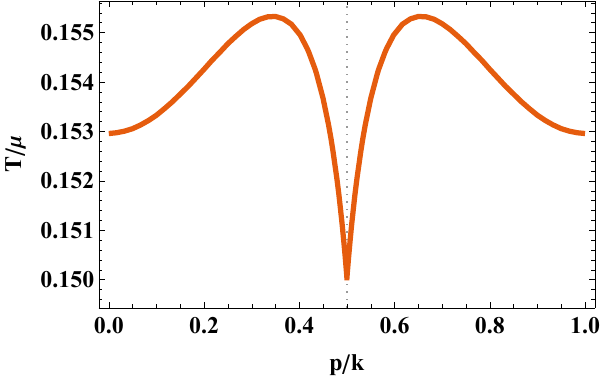}\end{minipage}\hfill
\begin{minipage}{0.32\textwidth}\centering
\includegraphics[width=\textwidth]{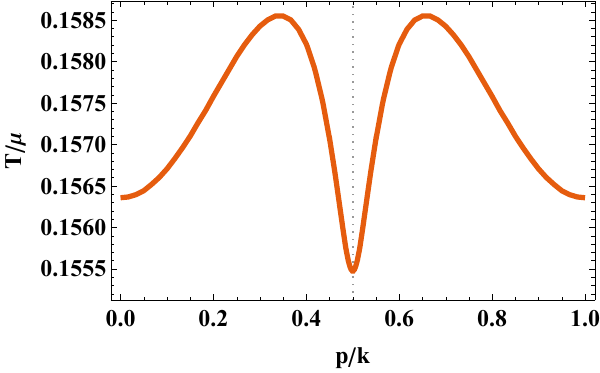}\end{minipage}\hfill
\begin{minipage}{0.32\textwidth}\centering
\includegraphics[width=\textwidth]{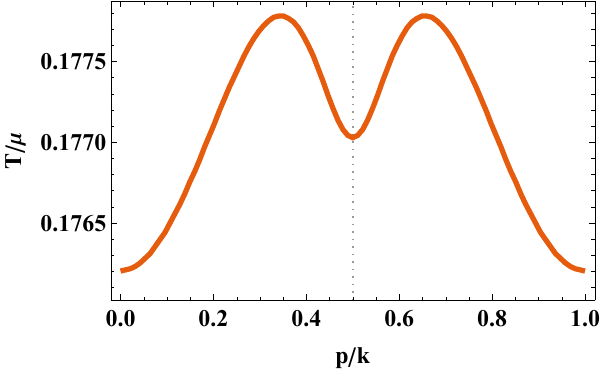}\end{minipage}

\vspace{0.1cm}

\begin{minipage}{0.32\textwidth}\centering
\includegraphics[width=\textwidth]{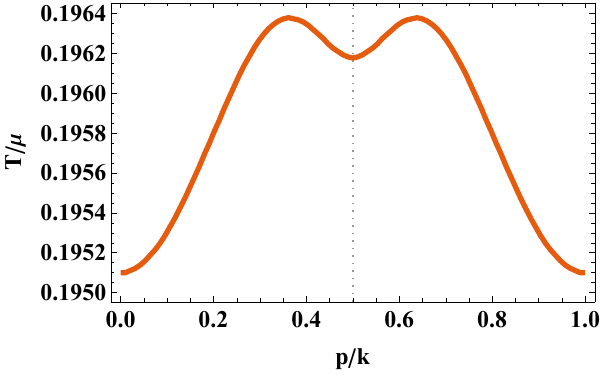}\end{minipage}\hfill
\begin{minipage}{0.32\textwidth}\centering
\includegraphics[width=\textwidth]{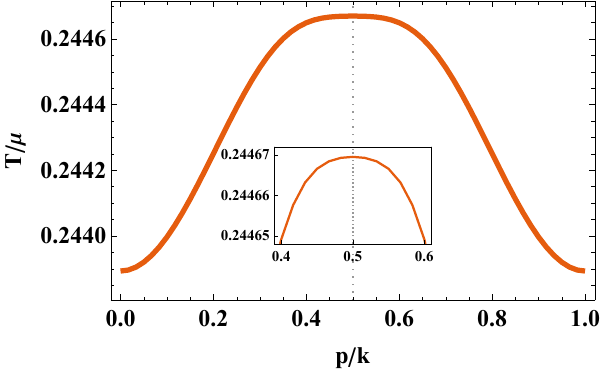}\end{minipage}\hfill
\begin{minipage}{0.32\textwidth}\centering
\includegraphics[width=\textwidth]{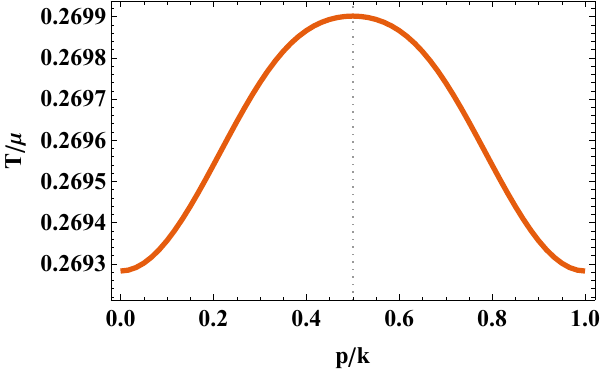}\end{minipage}
\caption{The lock-in process of the phase transition from $\mathrm{SC}$ to $\mathrm{SSC}$ in the $p$-$T$ plane at $e=3.9$. Each panel shows the critical temperature $T_c(p)$ as a function of the reduced wave-vector $p/k$. From top left to bottom right, the lattice amplitude is increased as $\lambda = 0, 0.4, 1, 1.3, 1.8, 2$. As $\lambda$ grows the maximum of $T_c(p)$ shifts to $p=k/2$ and finally locks there at $\lambda\approx1.8$. The inset in the panel with $\lambda=1.8$ magnifies the vicinity of $p=k/2$, confirming that the maximum indeed sits at the commensurate point.}
\label{fig:pT_lockin}
\end{figure}

In this subsection we demonstrate a novel phenomenon which is observed for the first time in the context of the holographic framework, that is, the commensurate lock-in effect in a striped superconducting phase. Previously the lock-in effect was observed only in an insulating phase with CDW in holographic setups~\cite{Andrade:2017leb,Krikun:2017cyw,Ling:2023ncu}. As a matter of fact, commensurate lock-in of density-wave and stripe orders is a familiar phenomenon in condensed-matter systems, reflecting the competition between the intrinsic wave-vector of a modulation and the pinning tendency supplied by the lattice~\cite{Gruner1988,McMillan1976,Bak1982}. Related commensurability effects also appear in striped superconductors and pair-density-wave settings~\cite{Berg_2009,Agterberg_2020}. In these standard descriptions, this competition can be controlled by the strength of a periodic substrate or Umklapp locking term; in the present holographic setup the ionic-lattice amplitude $\lambda$ plays this role. Therefore, we compute the critical temperature $T_c(p)$ for the instability of the superconducting background toward the SSC phase at different wave-vectors $p$ with different values of the lattice amplitude $\lambda$, and then plot the phase diagram on the $p$-$T$ plane, which is illustrated in Fig.~\ref{fig:pT_lockin}.

In this figure each plot illustrates the critical temperature $T_c(p)$ for the instability of the SC phase toward the SSC phase with a fixed lattice amplitude $\lambda$. It is manifestly observed that as $\lambda$ grows, the wave-vector with the highest $T_c$ migrates toward the half-harmonic position with $p=k/2$ and finally locks there, forming the commensurate $1{:}2$ state. Furthermore, it is interesting to compare the lock-in effect in the SSC phase observed here with the one observed in the CDW phase, which is illustrated in detail in~\cite{Ling:2023ncu}. First, with the same values of the parameters, the CDW locks at a relatively small lattice amplitude, $\lambda\approx 0.6$~\cite{Ling:2023ncu}, whereas the SSC mode locks at a substantially larger one, $\lambda\approx1.8$ (of course this number depends on the value of the charge $e$ as well). Second, the shape of the $T_c(p)$ curve changes more dramatically for the SC-to-SSC phase transition: the eventual lock-in point $p=k/2$ is in fact a sharp, narrow minimum of $T_c$ at small lattice amplitude, the bottom of a valley, but as $\lambda$ grows, the bottom of the valley rises fastest and finally becomes the tip of the curve, signaling the highest $T_c$ of the most stable mode. By contrast, the lock-in process for the CDW is illustrated in~\cite{Ling:2023ncu}, where the same point $p=k/2$ is merely a shallow dip on the top of a broad dome at $\lambda=0$; with the increase of $\lambda$, two tips beside it move toward the center and the dip becomes flatter, so that $p=k/2$ eventually turns into the tip of the curve and locks there in a much milder manner.

The above investigation of the phase diagrams can be performed at the linear perturbative level. Next we need to find the background solutions with backreaction by solving the equations of motion numerically, which involves a system of second-order partial differential equations. We apply the standard DeTurck method~\cite{Figueras_2011,Adam_2012} to find such static backgrounds with the CDW, SC and SSC orders separately, and briefly describe the main features of the order parameters and charge densities in Appendix~\ref{app:profiles}. We also compare the free energy of the coexisting branches at the same temperature and find that the SSC branch is thermodynamically preferred over the corresponding CDW or SC branch.

\section{The Optical Conductivity}\label{sec:sigma}
In this section we focus on the transport properties of the system by computing the optical conductivity using linear response theory. The most important task is to justify that the SSC background with nonzero order parameters $\eta$ and $\Phi$ indeed exhibits superconducting behavior. The optical conductivity of a system with a sole SC or CDW order has been extensively studied in the holographic literature. For a holographic superconductor where a nonzero condensate $\eta$ signals the breaking of the $U(1)$ symmetry, an infinite DC conductivity appears as a delta function at $\omega=0$ in $\mathrm{Re}\,\sigma$, which is indirectly observed through the Kramers--Kronig relation as a $1/\omega$ pole in $\mathrm{Im}\,\sigma$~\cite{Hartnoll:2008vx,Hartnoll:2008kx}. In addition, a hard gap is manifestly observed in the real part of the conductivity. For a holographic CDW where a nonzero scalar $\Phi$ signals the spontaneous breaking of translational invariance, by contrast, a nearly zero DC conductivity is observed, justifying its insulating nature. In addition, a pseudogap as well as the pinning effect is observed in the finite-frequency region of the real part of the conductivity~\cite{Ling:2014saa}. Now, the SSC phase, as the interplay of the CDW and SC orders, is believed to form a PDW and exhibit superconducting behavior, but due to the complexity of the numerical simulation, its optical conductivity has not been investigated in the literature. Next we intend to provide an affirmative answer to this question by computing the optical conductivity of a striped superconductor. In particular, we will demonstrate its transport behavior near the critical temperatures of the phase transitions, thus revealing its insulating nature in the CDW phase but superconducting nature in the SSC phase.

As described in Section~\ref{sec:phase}, we need to consider two distinct pathways into the SSC phase. For small charge $e$, the CDW forms first and superconductivity subsequently develops on an already modulated background; this tests how the superfluid response emerges from a pinned CDW state. For large charge $e$, the SC forms first and the SSC phase appears at lower temperature with PDW order; this tests how a spontaneous stripe modifies an existing superconducting transport channel. Furthermore, we are concerned with the impact of the lattice on the transport properties of the system. In the unlocked regime the explicit lattice provides only weak momentum relaxation and leaves the spontaneous stripe weakly pinned, so changes in the conductivity can be attributed more directly to the onset of whichever order appears second. The locked regime tests the opposite limit, in which the lattice already provides strong momentum relaxation and pins the modulation; there the key question is whether the additional spontaneous stripe still leaves an independent transport signature. Accordingly, we would like to investigate the optical conductivity of the system in two subsections. One is on the weak superconducting-pairing pathway (CDW $\to$ SSC), represented by $e=3.4$, while the other is on the strong superconducting-pairing pathway (SC $\to$ SSC), represented by $e=3.9$. In each subsection, we will consider the unlocked case and the locked case separately, with a focus on the conductivity as a function of $\omega/\mu$ near the phase transition close to $T_{\mathrm{SSC}}$.

Before diving into the numerical results of the optical conductivity, we briefly outline the procedure of the computation based on linear response theory. On a given inhomogeneous background, we perturb all fields as
\begin{equation}
\begin{aligned}
g_{\mu\nu} & =\bar g_{\mu\nu}+\delta g_{\mu\nu}, \quad
A_\mu=\bar A_\mu+\delta A_\mu, \quad
B_\mu=\bar B_\mu+\delta B_\mu, \\ 
\Phi & =\bar\Phi+\delta\Phi, \quad \quad\quad
\eta=\bar\eta+\delta\eta, \quad \quad\quad
\theta = \delta \theta,
\end{aligned}
\end{equation}
with all perturbations carrying the time dependence $e^{-i\omega t}$. We stress that although the phase $\theta$ vanishes on the background, its perturbation $\delta\theta$ must be turned on in linear response theory for consistency. To extract the optical conductivity along the lattice direction, we impose a nonzero boundary source for $\delta A_x$, and solve the full coupled system of linearized PDEs. We impose ingoing boundary conditions at the horizon and apply the standard de Donder and Lorenz gauge conditions for the metric and gauge-field perturbations, respectively. The detailed analysis on the computation of the optical conductivity over the holographic lattice can be found in Refs.~\cite{Horowitz:2012ky,Horowitz:2013jaa,Ling:2019gjy,Ling:2023ncu}.

Near the AdS boundary, the sourced perturbation has the Taylor expansion
\begin{equation}
\delta A_x(z,x,t)=\left[1+j_x(x)\,z+\mathcal{O}(z^2)\right]e^{-i\omega t}.
\end{equation}
According to Ohm's law, the optical conductivity is obtained as
\begin{equation}
\sigma(\omega)=\frac{4j_x^{(0)}}{i\omega},
\end{equation}
where $j_x^{(0)}$ is the uniform Fourier component of $j_x(x)$.

\subsection{Weak superconducting-pairing pathway ($e=3.4$): CDW $\to$ SSC}\label{subsec:sigma_e34}
We first consider the pathway of CDW $\to$ SSC. As shown in the $e$-$T$ phase diagram in Fig.~\ref{fig:four_phase_diagrams}, for $e=3.4$ the CDW appears first upon cooling, and the superconducting order appears only below $T_{\mathrm{SSC}}$. This realizes a situation where superconductivity develops on top of an already formed CDW background, a setting closely related to the phenomenology of underdoped cuprates, where superconductivity competes and coexists with charge order in the pseudogap regime~\cite{Berg_2009}. It should be stressed that the SSC order which emerges at $T_{\mathrm{SSC}}$ contains not only the uniform condensate $\eta_2^{(0)}$, as in the homogeneous holographic superconductor~\cite{Hartnoll:2008vx}, but also a modulated part $\eta_2^{(1)}$. More importantly, on this background the modulation of the condensate is imprinted not merely by the ionic lattice but also by the spontaneously formed CDW itself (Appendix~\ref{app:profiles}); the latter contribution constitutes a PDW component~\cite{Agterberg_2020}, so that the modulation of the condensate forms a striped superconductor rather than a purely uniform one. The pure CDW optical response on an ionic lattice was analyzed in Ref.~\cite{Ling:2023ncu}: after the CDW forms, the system undergoes a phase transition from the metallic phase to the insulating phase, as indicated by the suppression of the DC conductivity. The peak of the conductivity is shifted to a finite pinning frequency at lower temperature, and the suppression is stronger in the locked regime. Our focus here is therefore on whether the SSC transition generates a zero-frequency superfluid component, and whether it redistributes the regular optical spectral weight.

\begin{figure}[htp]
\centering
\begin{minipage}{0.49\textwidth}
\centering
\includegraphics[width=\textwidth]{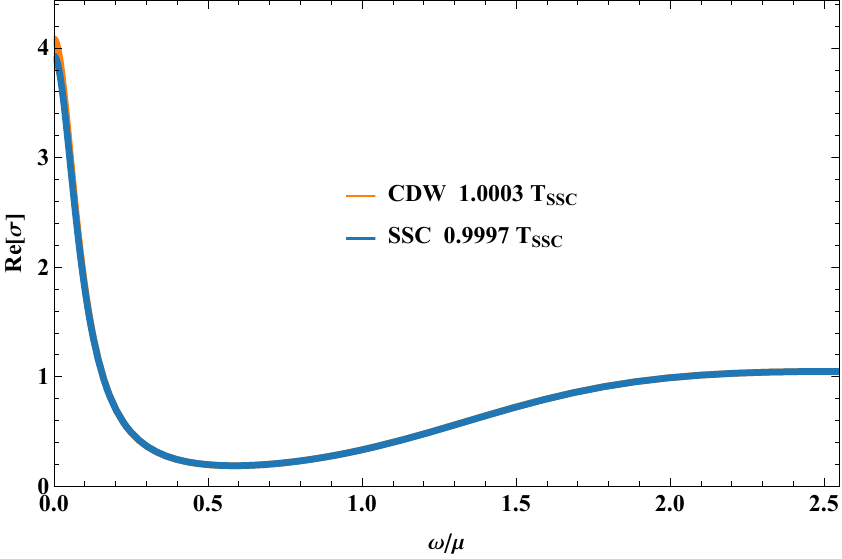}
\end{minipage}\hfill
\begin{minipage}{0.49\textwidth}
\centering
\includegraphics[width=\textwidth]{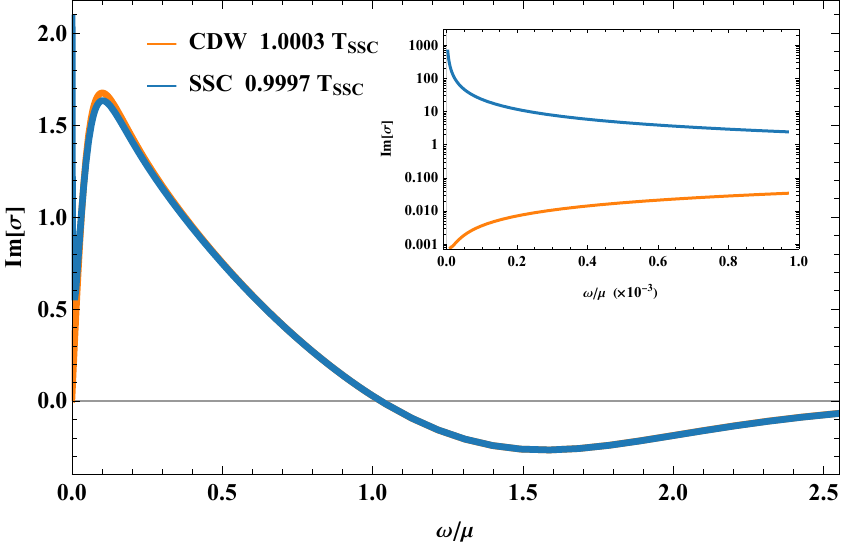}
\end{minipage}
\caption{The optical conductivity as a function of the frequency along the CDW-to-SSC pathway in the unlocked regime with $e=3.4$ and $\lambda=0.4$. The left panel is for the real part of the conductivity $\mathrm{Re}\,\sigma(\omega/\mu)$, while the right panel is for its imaginary part $\mathrm{Im}\,\sigma(\omega/\mu)$. In each panel two curves demonstrate the conductivity just above and just below the critical temperature $T_{\mathrm{SSC}}\approx0.1410\,\mu$, as indicated in the figure. The inset in the right panel zooms in on the region $\omega/\mu\le10^{-3}$ on a logarithmic vertical scale, where the SSC branch develops the $1/\omega$ superfluid pole while the CDW branch falls toward zero.}
\label{fig:sigma_e34}
\end{figure}

Fig.~\ref{fig:sigma_e34} demonstrates the optical conductivity of the system from CDW phase to SSC phase in the unlocked regime, $\lambda=0.4$, as a function of frequency. In this case the most important task is to justify that the system indeed undergoes a transition from an insulating phase to a superconducting phase. For this purpose we plot the optical conductivity at $T= 1.0003 \, T_{\mathrm{SSC}}$ (orange curve) and $T= 0.9997 \, T_{\mathrm{SSC}}$ (blue curve). The imaginary part identifies the superconducting response more directly. It is manifestly observed that in the CDW phase at $T= 1.0003 \, T_{\mathrm{SSC}}$, $\mathrm{Im}\,\sigma$ approaches zero in the zero-frequency limit, whereas in the SSC phase at $T= 0.9997 \, T_{\mathrm{SSC}}$, it becomes divergent, clearly indicating that a $1/\omega$ pole develops at $\omega=0$. By the Kramers--Kronig relation, this pole implies a delta function at $\omega=0$ in $\mathrm{Re}\,\sigma$, namely the superfluid weight. This confirms that the SSC phase is indeed superconducting.

Another feature along the CDW $\to$ SSC pathway is the formation of the pseudogap above $T_{\mathrm{SSC}}$. Since the CDW exhibits insulating behavior, it has already strongly suppressed the low-frequency normal transport. Low-frequency optical depletion associated with stripe correlations has also been observed above the bulk superconducting transition in La$_{2-x}$Ba$_x$CuO$_4$ at $x=1/8$~\cite{Homes2006LBCO,Homes2012LBCO}. The position of the pseudogap can be identified with the minimum of the imaginary part of the conductivity, located at $\omega \approx 1.6 \mu \approx 11.3 T_{\mathrm{SSC}}$. After entering the SSC phase, the regular real part\footnote{Throughout this paper, the regular part denotes the remainder after removing the zero-frequency superfluid contribution. We write $\mathrm{Re}\,\sigma(\omega)=\pi K_s\delta(\omega)+\mathrm{Re}\,\sigma_{\mathrm{reg}}(\omega)$ and $\mathrm{Im}\,\sigma(\omega)=K_s/\omega+\mathrm{Im}\,\sigma_{\mathrm{reg}}(\omega)$ for $\omega>0$, where $K_s$ is the superfluid weight. The imaginary part of the conductivity in the figures shows the full response, including the superfluid pole when present.} of the conductivity and the pseudogap-related optical features remain nearly unchanged, indicating that the gap-related optical scale and its strong-coupling character are inherited from the pre-existing CDW state rather than generated anew at $T_{\mathrm{SSC}}$.

\begin{figure}[htp]
\centering
\begin{minipage}{0.49\textwidth}
\centering
\includegraphics[width=\textwidth]{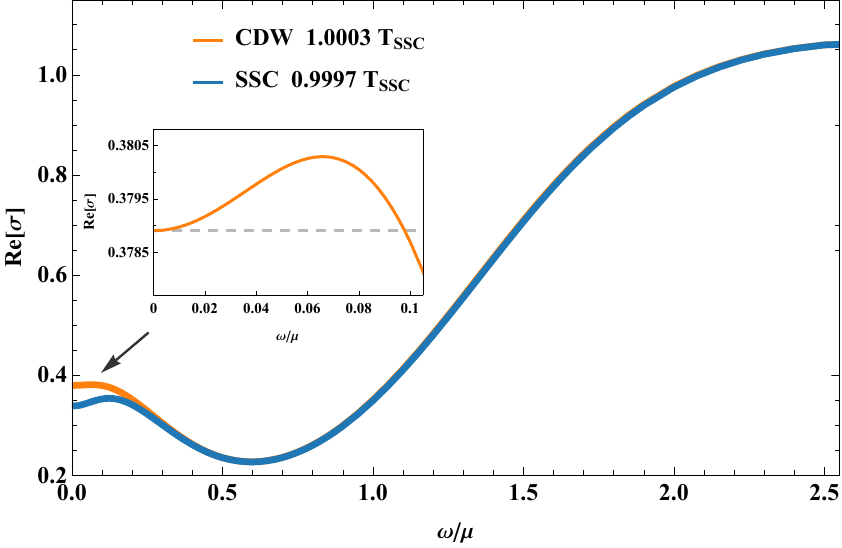}
\end{minipage}\hfill
\begin{minipage}{0.49\textwidth}
\centering
\includegraphics[width=\textwidth]{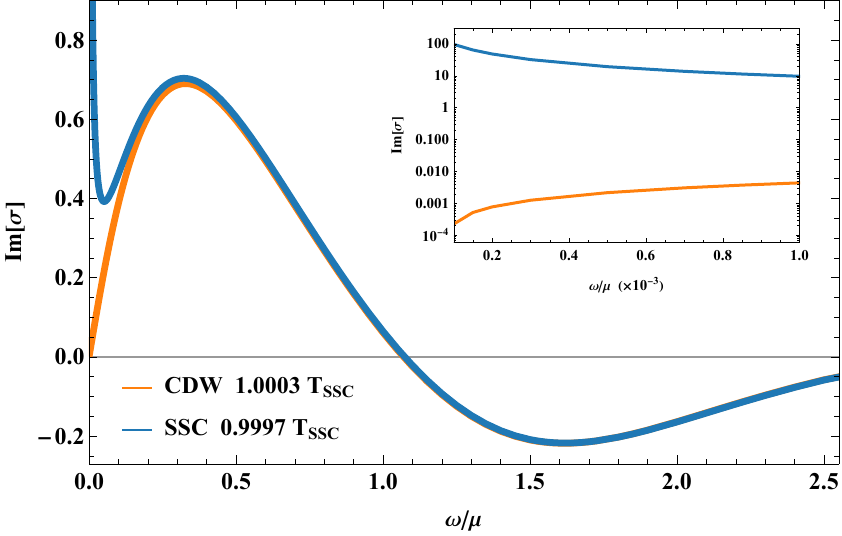}
\end{minipage}
\caption{The optical conductivity as a function of the frequency along the CDW-to-SSC pathway in the locked regime with $e=3.4$ and $\lambda=1.3$. The left panel is for the real part of the conductivity $\mathrm{Re}\,\sigma(\omega/\mu)$, while the right panel is for its imaginary part $\mathrm{Im}\,\sigma(\omega/\mu)$. In each panel two curves demonstrate the conductivity just above and just below the critical temperature $T_{\mathrm{SSC}}\approx0.1613\,\mu$, as indicated in the figure. The inset in the left panel zooms in on the near-DC region of the CDW branch, where the pinning effect occurs at $\omega/\mu\approx0.07$. The inset in the right panel zooms in on the region $\omega/\mu\le10^{-3}$ on a logarithmic vertical scale, where the SSC branch develops the $1/\omega$ superfluid pole while the CDW branch falls toward zero.}
\label{fig:sigma_e34_locked}
\end{figure}

As a comparison, we plot the optical conductivity of the system for the lock-in case in Fig.~\ref{fig:sigma_e34_locked}, where the lattice amplitude is set at $\lambda=1.3$. Firstly, we remark that the critical temperature for the phase transition becomes higher with the increase of the lattice amplitude, as already observed in the $e$-$T$ phase diagrams in Fig.~\ref{fig:four_phase_diagrams}. Now $T_{\mathrm{SSC}}\approx 0.1613\,\mu$, larger than $T_{\mathrm{SSC}}\approx 0.1410\,\mu$ with $\lambda=0.4$. Furthermore, in contrast to the unlocked case, we notice that the CDW suppresses the low-frequency normal transport far more strongly: the near-DC value of $\mathrm{Re}\,\sigma$ is smaller by about an order of magnitude, so that the pseudogap becomes more evident. More importantly, in the locked case ($\lambda=1.3$), the CDW phase exhibits a pinning feature, with a low-frequency maximum of $\mathrm{Re}\,\sigma$ at $\omega/\mu\approx0.07$ (see the inset of Fig.~\ref{fig:sigma_e34_locked}). In the SSC phase, the corresponding maximum occurs at $\omega/\mu\approx0.12$, as seen in the main panel. In contrast, no such finite-frequency peak appears in either phase in the unlocked case ($\lambda=0.4$) at the near-critical temperatures as shown in Fig.~\ref{fig:sigma_e34}; the regular real part of the conductivity is largest near $\omega=0$. Finite-frequency electronic modes interpreted as pinned CDW excitations have also been observed in the in-plane infrared response of underdoped YBa$_2$Cu$_3$O$_{6+x}$ (YBCO), including below the superconducting transition~\cite{Lyzwa2020YBCO}. It is also interesting to notice that a similar phenomenon is observed in La$_{1.8-x}$Eu$_{0.2}$Sr$_x$CuO$_4$ (LESCO) at $x=1/8$, where the pre-existing far-infrared peak shifts to higher frequencies upon cooling below the charge-ordering temperature, while the low-frequency optical conductivity is further suppressed~\cite{Autore2014LESCO}. Moreover, since the low-frequency $\mathrm{Re}\,\sigma$ is already strongly suppressed at $T_{\mathrm{SSC}}$, upon further cooling the hard gap forms more rapidly than in the unlocked case. Away from this low-frequency region, the conductivity in the SSC phase and in the CDW phase behaves very similarly. The imaginary part again displays the superconducting signature: the SSC curve develops the $1/\omega$ pole associated with the zero-frequency superfluid weight. Thus we conclude that in the lock-in state the phase transition from CDW $\to$ SSC becomes easier, as the critical temperature $T_{\mathrm{SSC}}$ becomes higher. Furthermore, the formation of the pseudogap with evident pinning effects benefits the formation of the hard gap in SSC phase. Such phenomena are consistent with the experimental observations in the cuprates, where the superconducting gap emerges from the normal-state pseudogap~\cite{Timusk_1999,Kanigel_2006}.

\begin{figure}[htp]
\centering
\begin{minipage}{0.49\textwidth}
\centering
\includegraphics[width=\textwidth]{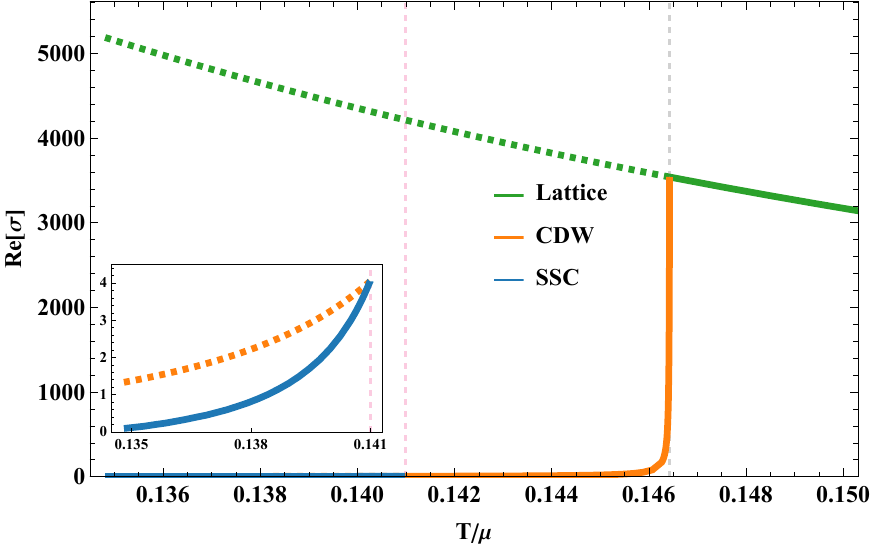}
\end{minipage}\hfill
\begin{minipage}{0.49\textwidth}
\centering
\includegraphics[width=\textwidth]{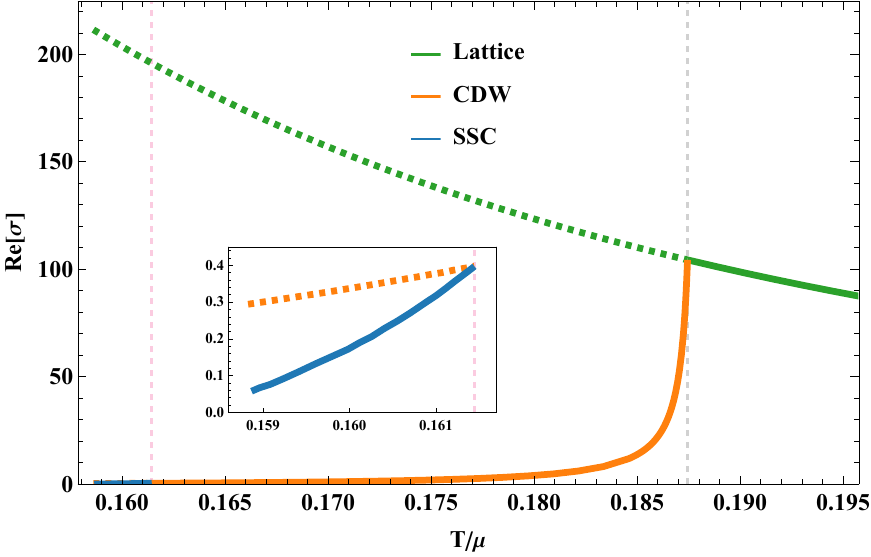}
\end{minipage}
\caption{Near-DC real conductivity $\mathrm{Re}\,\sigma$ at $\omega/\mu=10^{-10}$ as a function of temperature along the CDW-to-SSC pathway ($e=3.4$). The left panel is for the unlocked case with $\lambda=0.4$ and the right panel is for the locked case with $\lambda=1.3$. The vertical dashed lines mark $T_{\mathrm{CDW}}$ (gray) and $T_{\mathrm{SSC}}$ (pink). Solid lines denote the thermodynamically preferred branch, and the remaining dashed lines denote metastable branches.}
\label{fig:sigma_e34_varT}
\end{figure}

To understand the effects of CDW on the change of the conductivity over the whole course of the phase transitions, it is helpful to plot the real part of the near-DC conductivity as a function of temperature, as illustrated in Fig.~\ref{fig:sigma_e34_varT}, where $\mathrm{Re}\,\sigma$ is the value at a tiny but finite frequency $\omega/\mu=10^{-10}$, that is, the regular low-frequency response rather than the superconducting delta function itself. First of all, we notice that above $T_{\mathrm{CDW}}$ the system lies in the metallic phase and the presence of the ionic lattice makes it exhibit standard Drude behavior so that the near-DC conductivity is correspondingly large, with a magnitude set by the lattice amplitude. Below $T_{\mathrm{CDW}}$, the onset of the CDW collapses it rapidly, by more than two orders of magnitude on the way down to $T_{\mathrm{SSC}}$ in both regimes---the metal-to-insulator transition already identified in the purely CDW version of this model~\cite{Ling:2023ncu}. Below $T_{\mathrm{SSC}}$ the near-DC conductivity continues to fall without any abrupt change, though faster than along the metastable CDW branch (see the insets). The regular near-DC channel is thus governed throughout by the modulated background: the collapse happens at $T_{\mathrm{CDW}}$, and crossing $T_{\mathrm{SSC}}$ produces nothing comparable to it. In addition, the near-DC conductivity of the system exhibits similar behavior in both the unlocked and locked cases, except that the magnitude of the near-DC conductivity in the locked case becomes much smaller than that in the unlocked case, as seen by comparing the two panels of Fig.~\ref{fig:sigma_e34_varT}.

\begin{figure}[htp]
\centering
\begin{minipage}{0.49\textwidth}
\centering
\includegraphics[width=\textwidth]{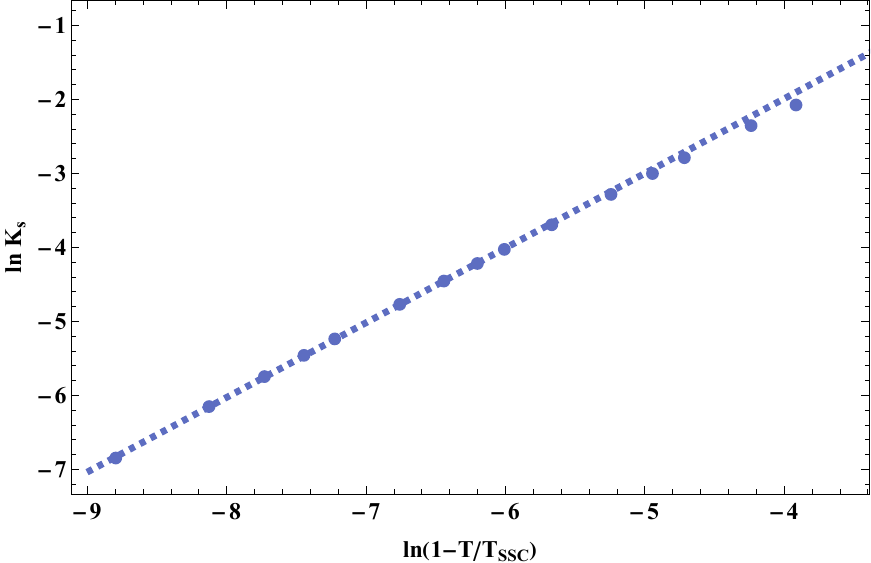}
\end{minipage}\hfill
\begin{minipage}{0.49\textwidth}
\centering
\includegraphics[width=\textwidth]{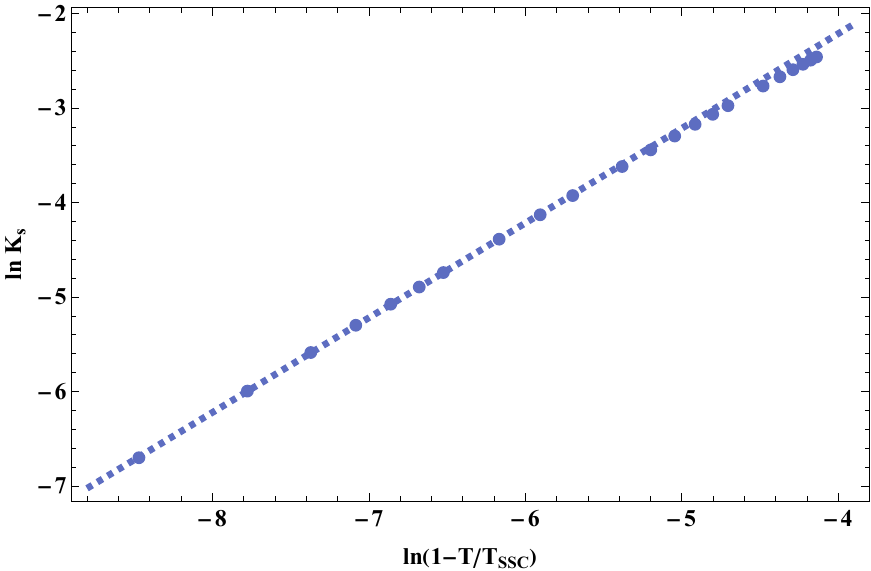}
\end{minipage}
\caption{The superfluid weight $K_s$ as a function of the reduced temperature $\delta T=1-T/T_{\mathrm{SSC}}$ for $e=3.4$, shown on logarithmic scales. The left panel is for the unlocked case with $\lambda=0.4$, while the right panel is for the locked case with $\lambda=1.3$. Points denote numerical data, and dashed lines are power-law fits, $K_s\propto(\delta T)^\alpha$, over the near-critical range $\ln\delta T\lesssim-5.9$. The fitted exponents are $\alpha=1.009$ and $\alpha=1.002$, respectively. In both regimes, $K_s$ vanishes approximately linearly as $T\to T_{\mathrm{SSC}}$, consistent with the mean-field scaling $K_s\propto\eta^2$ with $\eta\propto(\delta T)^{1/2}$.}
\label{fig:Ks_e34}
\end{figure}

We next quantify the superfluid response by extracting its weight $K_s$ from the imaginary part of the conductivity in the zero-frequency limit, namely
\begin{equation}
K_s(T)=\lim_{\omega\to0}\omega\,\mathrm{Im}\,\sigma(\omega,T),
\end{equation}
and fit its dependence on the reduced temperature $\delta T=1-T/T_{\mathrm{SSC}}$ to a power law,
\begin{equation}
K_s=A(\delta T)^\alpha ,
\end{equation}
which gives $\alpha=1.009$ for $\lambda=0.4$ and $\alpha=1.002$ for $\lambda=1.3$ in the near-critical window (Fig.~\ref{fig:Ks_e34}). The superfluid weight therefore vanishes linearly in $\delta T$, the mean-field behavior expected from $K_s\propto\eta^2$ with $\eta\propto(\delta T)^{1/2}$, as in the standard holographic superconductor~\cite{Hartnoll:2008vx}.

\subsection{Strong superconducting-pairing pathway ($e=3.9$): SC $\to$ SSC}\label{subsec:sigma_e39}
We next turn to investigate the optical conductivity for the pathway of SC $\to$ SSC. As shown in the $e$-$T$ phase diagram in Fig.~\ref{fig:four_phase_diagrams}, for $e>e_\ast$ the system first enters the SC phase upon cooling since the critical temperature $T_{\mathrm{SC}}$ for $U(1)$ symmetry breaking is higher than the critical temperature $T_{\mathrm{CDW}}$ for translational symmetry breaking. Subsequently, it directly enters the SSC phase below $T_{\mathrm{SSC}}$, where the spontaneous stripe appears. It should be stressed that this stripe is not merely a CDW: the condensate, so far modulated only by the imprint of the ionic lattice, now acquires a spontaneous modulation of its own, and it is this PDW component that turns the SC into a striped superconductor~\cite{Berg_2009,Agterberg_2020}. This is the complementary situation to the previous subsection: the stripe now develops on a background that is already superconducting. Our focus here is therefore on whether the superfluid pole survives after the onset of the stripe, and how the stripe modifies the existing superconducting transport channel. This should also be contrasted with the pure CDW case of Ref.~\cite{Ling:2023ncu} recalled in the previous subsection, where the CDW develops on the bare, non-superconducting ionic lattice.

\begin{figure}[htp]
\centering
\begin{minipage}{0.49\textwidth}
\centering
\includegraphics[width=\textwidth]{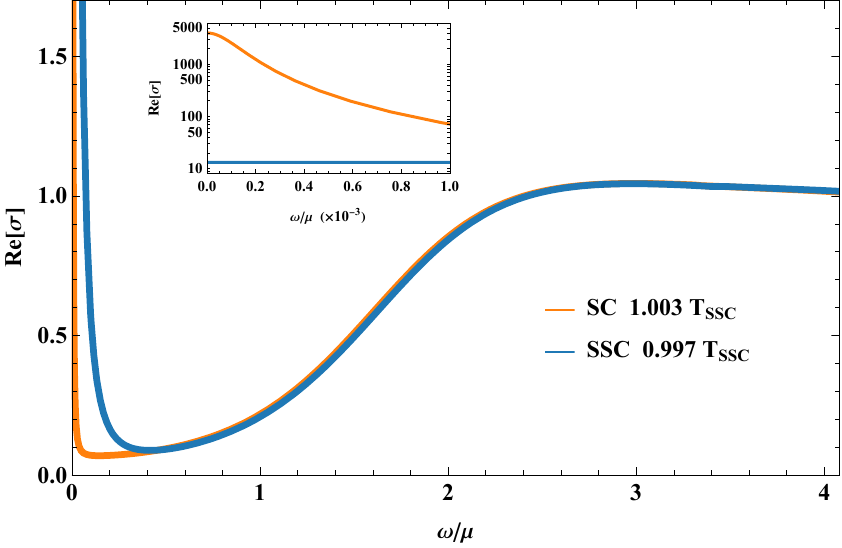}
\end{minipage}\hfill
\begin{minipage}{0.49\textwidth}
\centering
\includegraphics[width=\textwidth]{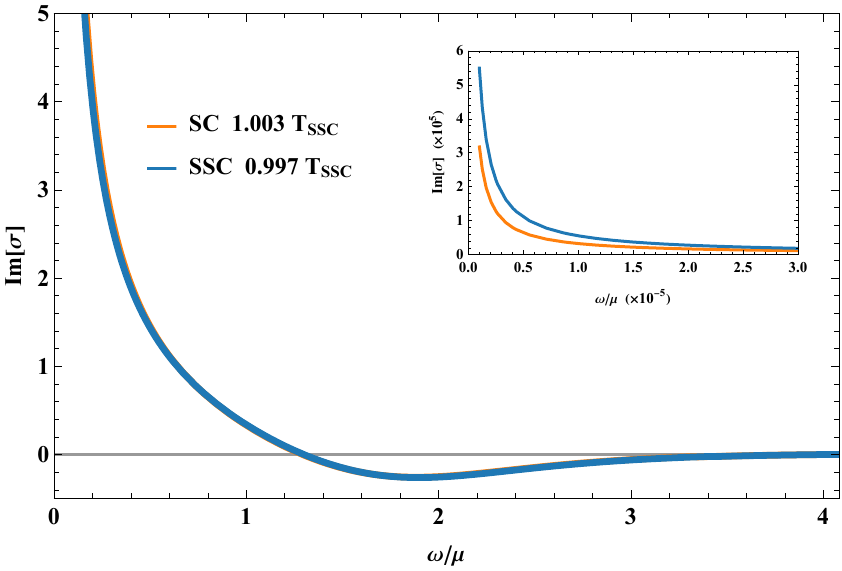}
\end{minipage}
\caption{The optical conductivity as a function of frequency along the SC-to-SSC pathway in the unlocked regime with $e=3.9$ and $\lambda=0.4$. The left panel is for the real part of the conductivity $\mathrm{Re}\,\sigma(\omega/\mu)$, while the right panel is for its imaginary part $\mathrm{Im}\,\sigma(\omega/\mu)$. In each panel two curves demonstrate the conductivity just above and just below the critical temperature $T_{\mathrm{SSC}}\approx0.1555\,\mu$, as indicated in the figure. The inset in the left panel zooms in on the region $\omega/\mu\le10^{-3}$ on a logarithmic vertical scale, showing the sharp near-DC peak in the SC phase, with a height of approximately $4\times10^3$, while the SSC curve remains nearly flat at approximately $13$. The inset in the right panel zooms in on the region $\omega/\mu\le3\times10^{-5}$, where both phases display the $1/\omega$ superfluid pole, with $K_s$ larger in the SSC phase than in the SC phase.}
\label{fig:sigma_e39}
\end{figure}

Fig.~\ref{fig:sigma_e39} demonstrates the optical conductivity of the system from SC phase to SSC phase in the unlocked regime, $\lambda=0.4$, as a function of frequency close to $T_{\mathrm{SSC}}$. In contrast to the previous pathway where the CDW forms prior to the SC, now the SC phase contains a very sharp near-DC peak in the regular part of $\mathrm{Re}\,\sigma$, visible in the inset, which reflects a highly coherent residual normal-fluid channel coexisting with the superfluid response. We remark that a narrow dissipative peak coexisting with the superfluid response is also observed in microwave measurements of YBCO~\cite{Hosseini1999Microwave,Harris2006Microwave}. After the stripe forms, the $1/\omega$ pole persists in the imaginary part of the conductivity $\mathrm{Im}\,\sigma$, clearly indicating that the striped state is indeed still a superconductor. In particular, $K_s$ is larger in the SSC phase than in the SC phase, implying that the superfluid component not only survives but is substantially enhanced. This enhancement of the superfluid response reflects cooperation between the CDW and SC orders, consistent with the phase diagrams in Sec.~\ref{subsec:eT}. Furthermore, from the inset of the left panel, we notice that the narrow near-DC peak is strongly suppressed in the presence of PDW, while in the main panel $\mathrm{Re}\,\sigma$ in the SSC state (the blue curve) becomes larger than that in the SC state (the orange curve) over a broader low-frequency region. This behavior indicates that the spontaneous stripe strongly modifies the residual normal-fluid channel, broadening and redistributing the regular low-frequency response rather than uniformly suppressing the conductivity.

\begin{figure}[htp]
\centering
\begin{minipage}{0.49\textwidth}
\centering
\includegraphics[width=\textwidth]{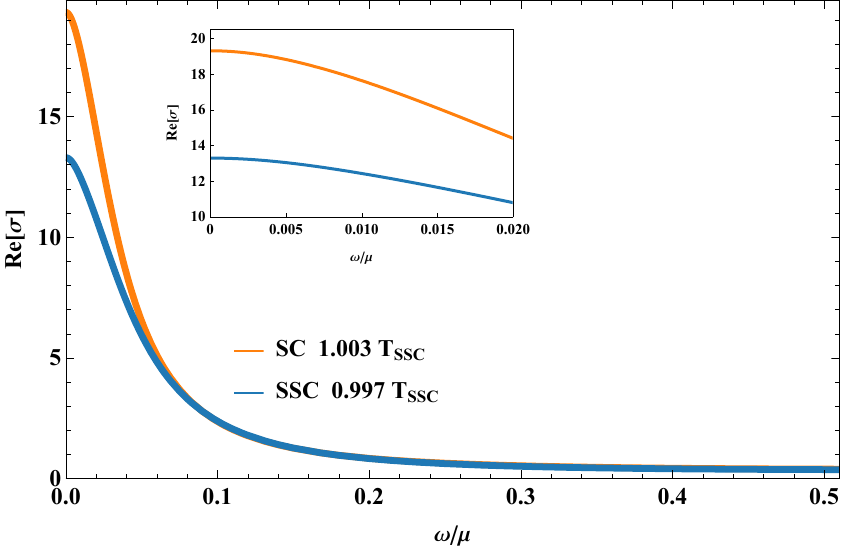}
\end{minipage}\hfill
\begin{minipage}{0.49\textwidth}
\centering
\includegraphics[width=\textwidth]{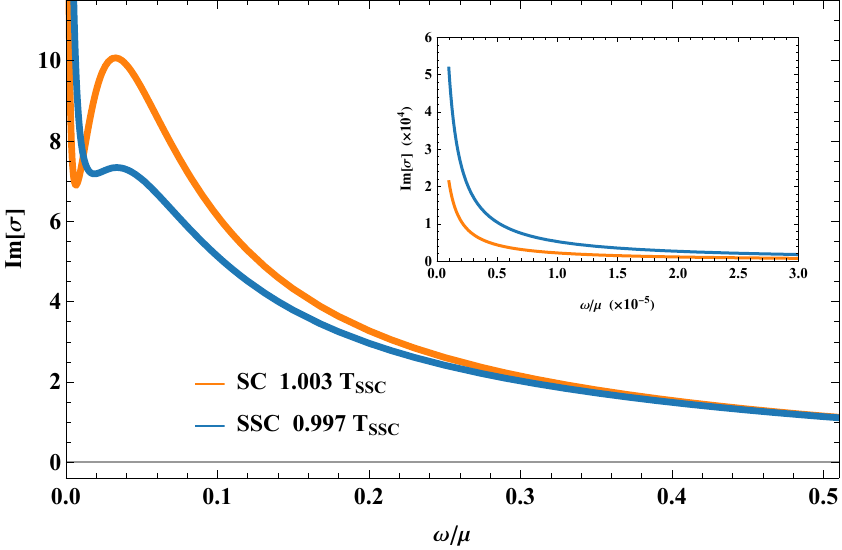}
\end{minipage}
\caption{The optical conductivity as a function of frequency along the SC-to-SSC pathway in the locked regime with $e=3.9$ and $\lambda=1.8$. The left panel is for the real part of the conductivity $\mathrm{Re}\,\sigma(\omega/\mu)$, while the right panel is for its imaginary part $\mathrm{Im}\,\sigma(\omega/\mu)$. In each panel two curves demonstrate the conductivity just above and just below the critical temperature $T_{\mathrm{SSC}}\approx0.2447\,\mu$, as indicated in the figure. The inset in the left panel zooms in on the region $\omega/\mu\le0.02$. The inset in the right panel zooms in on the region $\omega/\mu\le3\times10^{-5}$, where both phases display the $1/\omega$ superfluid pole, with $K_s$ larger in the SSC phase than in the SC phase.}
\label{fig:sigma_locked}
\end{figure}

Next we move to the lock-in case by setting $\lambda=1.8$, where the spontaneous modulation is commensurate with the ionic lattice at $p=k/2$ (see Section~\ref{subsec:pT}). As illustrated in Fig.~\ref{fig:sigma_locked}, the imaginary part shows the same behavior as in the unlocked case: both phases retain the $1/\omega$ superfluid pole, while the low-frequency inset indicates that $K_s$ is considerably larger in the SSC phase than in the SC phase. The difference lies in the real part. The SC phase no longer contains the enormous narrow peak with huge DC conductivity; instead, the regular part of $\mathrm{Re}\,\sigma$ is already reduced to the scale of $\mathcal{O}(10)$ and decreases smoothly with frequency. After the stripe appears, the SSC curve is further suppressed near $\omega=0$ but remains monotonically decreasing, so that no pinning maximum develops at finite frequency, in contrast to the locked CDW $\to$ SSC pathway (Fig.~\ref{fig:sigma_e34_locked}); at finite frequency it becomes practically indistinguishable from the SC curve. Thus the locked background has already removed most of the coherent residual normal-fluid channel before the SC-to-SSC transition takes place, and the additional stripe produces only a mild change in the regular conductivity. The main effect of the SSC transition in the locked regime is therefore not a broad reconstruction of the optical spectrum, but a modest suppression of the regular near-DC response together with a strengthened singular superfluid contribution.

\begin{figure}[htp]
\centering
\begin{minipage}{0.49\textwidth}
\centering
\includegraphics[width=\textwidth]{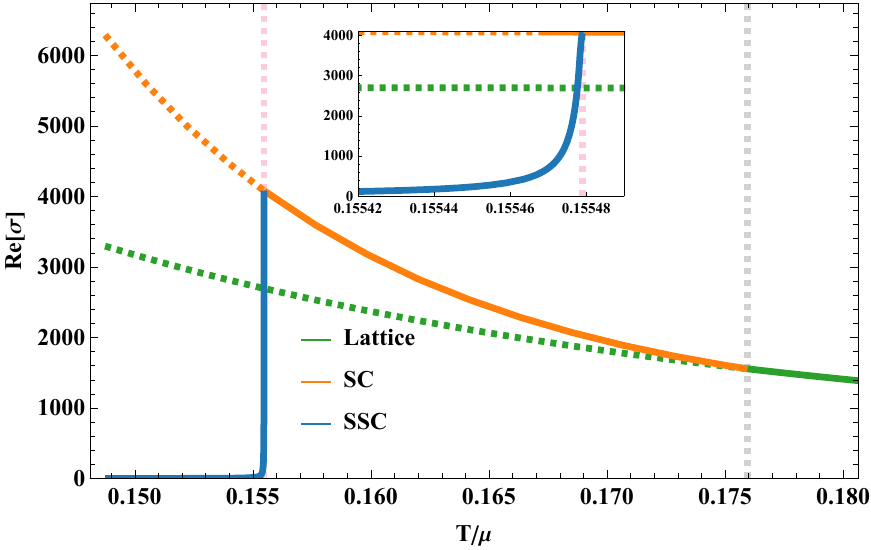}
\end{minipage}\hfill
\begin{minipage}{0.49\textwidth}
\centering
\includegraphics[width=\textwidth]{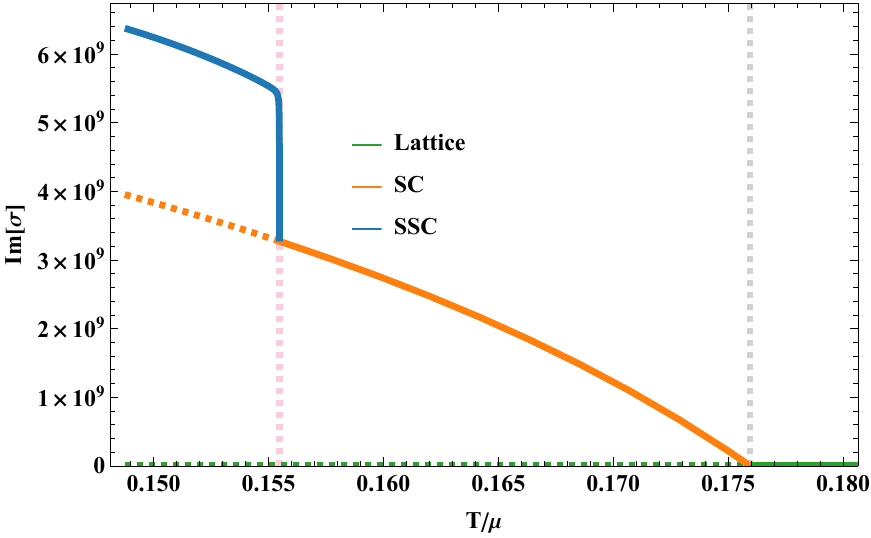}
\end{minipage}
\caption{Near-DC conductivity at $\omega/\mu=10^{-10}$ as a function of temperature along the SC-to-SSC pathway in the unlocked regime with $e=3.9$ and $\lambda=0.4$. The left panel is for the real part of the conductivity $\mathrm{Re}\,\sigma$, while the right panel is for its imaginary part $\mathrm{Im}\,\sigma$. The vertical dashed lines mark $T_{\mathrm{SC}}$ (gray) and $T_{\mathrm{SSC}}$ (pink). Solid lines denote the thermodynamically preferred branch, and the remaining dashed lines denote metastable branches.}

\label{fig:sigma_e39_varT}
\end{figure}

\begin{figure}[htp]
\centering
\begin{minipage}{0.49\textwidth}
\centering
\includegraphics[width=\textwidth]{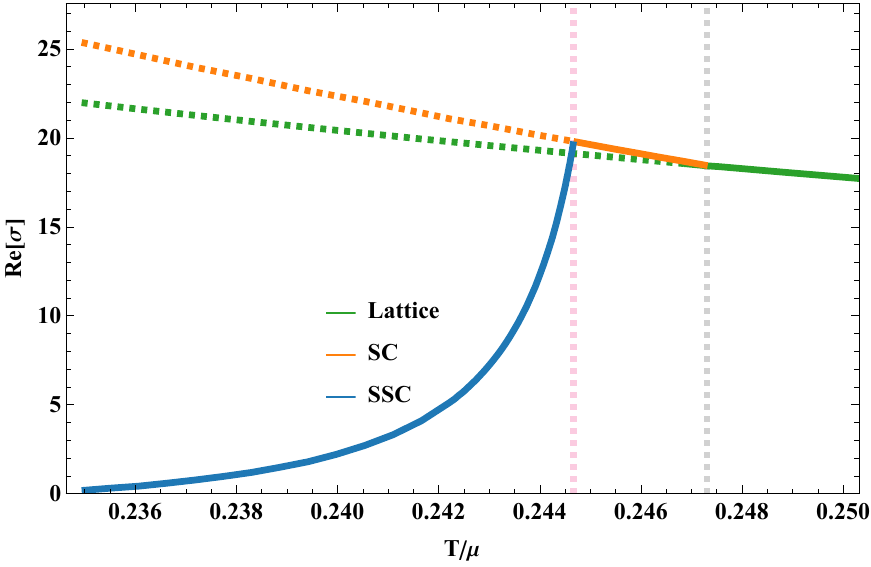}
\end{minipage}\hfill
\begin{minipage}{0.49\textwidth}
\centering
\includegraphics[width=\textwidth]{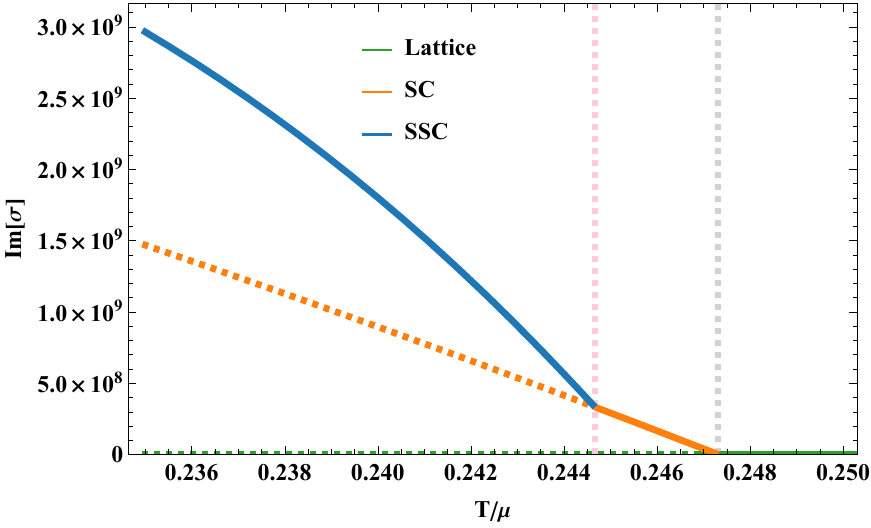}
\end{minipage}
\caption{Near-DC conductivity at $\omega/\mu=10^{-10}$ as a function of temperature along the SC-to-SSC pathway in the locked regime with $e=3.9$ and $\lambda=1.8$. The left panel is for the real part of the conductivity $\mathrm{Re}\,\sigma$, while the right panel is for its imaginary part $\mathrm{Im}\,\sigma$. The vertical dashed lines mark $T_{\mathrm{SC}}$ (gray) and $T_{\mathrm{SSC}}$ (pink). Solid lines denote the thermodynamically preferred branch, and the remaining dashed lines denote metastable branches.}

\label{fig:sigma_locked_dc}
\end{figure}

Figs.~\ref{fig:sigma_e39_varT} and \ref{fig:sigma_locked_dc} show the temperature dependence of the near-DC conductivity in the unlocked and locked regimes, respectively. The real part of the conductivity shows the regular response, while the imaginary part at this near-DC frequency tracks the superfluid weight. In the unlocked regime the real part of the conductivity in the SC phase grows rapidly upon cooling from $T_{\mathrm{SC}}$ and reaches its maximum at $T_{\mathrm{SSC}}$; its value is just the height of the coherent near-DC peak of Fig.~\ref{fig:sigma_e39}. It is remarkable that once the stripe appears below $T_{\mathrm{SSC}}$, this peak drops dramatically, and the drop is essentially complete within a fraction of a percent in temperature below $T_{\mathrm{SSC}}$ (as illustrated in the inset of Fig.~\ref{fig:sigma_e39_varT}). By contrast, in the locked regime a similar phenomenon is observed but at a far smaller scale: the real part of the conductivity stays at the scale of tens imposed by the strong lattice, and below $T_{\mathrm{SSC}}$ it decreases smoothly over several percent in temperature rather than dropping abruptly, as illustrated in Fig.~\ref{fig:sigma_locked_dc}. The imaginary part of the conductivity behaves alike in both regimes. It vanishes in the metallic phase with the ionic lattice, but turns on at $T_{\mathrm{SC}}$ on the SC branch and grows on cooling; below $T_{\mathrm{SSC}}$, on the SSC branch, it grows more steeply than on the metastable continuation of the SC branch---the enhancement of the superfluid pole seen in the frequency scans, now along the temperature axis. The temperature scans thus display two effects of the stripe side by side: the regular near-DC channel drops at $T_{\mathrm{SSC}}$, abruptly in the unlocked regime and gradually in the locked one, while the superfluid weight, already growing on the SC branch, grows faster still once the stripe is present.

\section{Conclusion and Discussion}\label{sec:conclusion}
In this paper we have constructed a holographic model of a striped superconductor characterized by the following two features. Firstly, the PDW order in the striped superconducting phase emerges from the coexistence of the CDW and SC orders. Secondly, this model is built on an explicit ionic lattice so as to exhibit a lock-in behavior, and the locked CDW phase may serve as a Mott-like parent state. Furthermore, in contrast to Ref.~\cite{Ling:2019gjy}, the model considered in this paper contains free charge carriers and thus is closer to the real striped superconductors observed in the laboratory, providing a concrete holographic understanding for the intertwined CDW, SC and PDW orders observed in the cuprates. We have analyzed its phase structures in detail with the standard perturbative methods over a fixed background with the ionic lattice. The phase diagram on the $e$-$T$ plane reveals that the SSC phase exists only within an intermediate window of $e$ and is reached through distinct phase transitions depending on the value of $e$: for small $e$, upon cooling the system first enters the CDW phase from the metallic phase, and subsequently the SSC phase once the $U(1)$ gauge symmetry is spontaneously broken; while for large $e$, the system enters the SC phase from the metallic phase directly, and subsequently the SSC phase once the translational symmetry is spontaneously broken. The cooperative relation between the CDW order and SC order is reflected in the fact that the critical temperature $T_{\mathrm{SSC}}$ always lies above the bare critical temperature of the order that appears second. On the other hand, the window itself reflects the competitive side: both orders are driven by the same gauge field $A_\mu$ and compete for the same carriers, and once the charge $e$ is large enough the CDW no longer forms so that the SSC phase is absent. The $p$-$T$ phase diagram, in turn, displays for the first time in holography the commensurate lock-in of a PDW order, which may be viewed as a significant extension of the commensurate lock-in effect observed in the CDW phase in previous holographic literature~\cite{Andrade:2017leb,Krikun:2017cyw,Ling:2023ncu}. More importantly, this lock-in mechanism may be relevant to the commensurability effects observed in striped superconductors and pair-density-wave systems~\cite{Berg_2009,Agterberg_2020}, and thus shed light on the pairing mechanism in high temperature superconductors in the future.

We have performed a systematic analysis of the optical conductivity of the holographic striped superconductor. For the first time in the context of holography we have explicitly demonstrated that the system in SSC phase indeed exhibits superconducting behavior, supported by the evidence that the imaginary part of the conductivity becomes divergent in the zero-frequency limit just below the critical temperature $T_{\mathrm{SSC}}$. The distinct behavior of the conductivity along different pathways to SSC phase is also investigated in detail. The pathway of CDW $\to$ SSC is more likely to be observed in real materials, where the system undergoes a transition from insulating phase to the superconducting phase, with the formation of the pseudogap and the pinning effects, which become more transparent in the lock-in case where the lattice effect is strong with a large lattice amplitude. The pathway of SC $\to$ SSC, by contrast, is more theoretically interesting, where the modulated modes are developed directly over an SC order as soon as the translational symmetry is spontaneously broken. It is found that the near-DC value of $\mathrm{Re}\,\sigma$ decreases after the PDW forms.

Our current work provides robust evidence for the superconductivity of the striped phase constructed in the literature of holography, and a lot of work may be pushed forward in the near future. We would like to list some crucial issues that should be addressed in the next investigation. Firstly, to focus on the optical conductivity, we ignore the doping effect throughout this paper. This could be introduced by turning on both gauge fields for the background in this model, as in Refs.~\cite{Ling:2020qdd,Li:2024ybq,Cai:2020nyd}. Secondly, the anisotropic behavior of the optical conductivity could be investigated, since in condensed matter physics many materials exhibit anisotropic conductivity along different spatial directions, and the striped superconductors are prominent examples~\cite{Berg_2009,Vojta01112009}. Thirdly, our current model may be viewed as an extension of the seminal work in Ref.~\cite{Andrade:2017ghg}. Some substantial progress has been made on the road of understanding the high-$T_c$ superconductors in the holographic approach. The Mott-like nature of the parent compounds of high-$T_c$ superconductors could be explored further based on our current model in the future. Finally, we have only focused on the behavior of the optical conductivity near the critical temperature $T_{\mathrm{SSC}}$. The lower-temperature behavior, including quantum criticality in the zero-temperature limit, may be investigated in the future.

\section*{Acknowledgments}
We are very grateful to Bin-ye Dong and Zhang-ping Yu for helpful discussions. This work is supported in part by the National Natural Science Foundation of China (Grant Nos.~12275275,~12675082,~12361141825,~12575047 and~12405067). 

\appendix
\section{Equations of motion}\label{app:eoms}
We rewrite the $U(1)$ charged complex scalar field as $\Psi=\eta\,e^{i\theta}$, where $\eta \geq 0$ is a real scalar field and $\theta$ is the phase of $\Psi$, so that the action in Eq.~\eqref{eq:action} can be cast as
\begin{equation}\label{eq:action_polar}
\begin{aligned}
S = & \frac{1}{2\kappa^2} \int d^4x \sqrt{-g} \bigg[ R - \frac{1}{2}(\nabla\Phi)^2 - V(\Phi) - \frac{1}{4} Z_A(\Phi)F^2 \\
    &- \frac{1}{4} Z_B(\Phi)G^2 - \frac{1}{2} Z_{AB}(\Phi)FG - (\nabla\eta)^2 - m_v^2\eta^2 - \eta^2\left(\nabla\theta - e A\right)^2 \bigg].
\end{aligned}
\end{equation}
Varying~\eqref{eq:action_polar} yields the equations of motion
\begin{equation}
\begin{aligned}
R_{\mu\nu} - T^{\Phi}_{\mu\nu} - T^{A}_{\mu\nu} - T^{B}_{\mu\nu} - T^{AB}_{\mu\nu} - T^{\eta}_{\mu\nu} - T^\theta_{\mu \nu} & = 0,\\
\nabla^2\Phi - \tfrac14 Z_A' F^2 - \tfrac14 Z_B' G^2 - \tfrac12 Z_{AB}' FG - V' & = 0,\\
\nabla^2\eta - m_v^2\eta - ( \nabla \theta - e A)^2 \eta & = 0,\\
\nabla_\mu\!\big(Z_A F^{\mu\nu} + Z_{AB} G^{\mu\nu}\big) + 2 e \eta^2 ( \nabla^\nu \theta - e A^\nu ) & = 0,\\
\nabla_\mu\!\big(Z_B G^{\mu\nu} + Z_{AB} F^{\mu\nu}\big) & = 0,\\
\nabla_\mu (\eta^2 (\nabla^\mu \theta - e A^\mu )) & = 0,
\end{aligned}
\end{equation}
where a prime denotes a derivative with respect to $\Phi$, and the stress-energy tensors are given by
\begin{equation}
\begin{aligned}
T^{\Phi}_{\mu\nu} &= \tfrac12\nabla_\mu\Phi\nabla_\nu\Phi + \tfrac12 V g_{\mu\nu},\\
T^{A}_{\mu\nu} &= \tfrac{Z_A}{2}\big(F_{\mu\rho}F_\nu{}^{\rho} - \tfrac14 g_{\mu\nu}F^2\big),\\
T^{B}_{\mu\nu} &= \tfrac{Z_B}{2}\big(G_{\mu\rho}G_\nu{}^{\rho} - \tfrac14 g_{\mu\nu}G^2\big),\\
T^{AB}_{\mu\nu} &= Z_{AB}\big(F_{(\mu|\rho|}G_{\nu)}{}^{\rho} - \tfrac14 g_{\mu\nu}FG\big),\\
T^{\eta}_{\mu\nu} &= \nabla_\mu\eta\nabla_\nu\eta + e^2\eta^2 A_\mu A_\nu + \tfrac12 m_v^2\eta^2 g_{\mu\nu},\\
T^\theta_{\mu \nu} & = \eta^2 ( \nabla_\mu \theta \nabla_\nu \theta - 2 e A_{(\mu} \nabla_{\nu )} \theta ).
\end{aligned}
\end{equation}

\section{Order-Parameter and Charge-Density Profiles}\label{app:profiles}
For completeness, this appendix presents the temperature dependence of the order parameters and charge densities. As discussed in the main text, the bulk fields have the following expansions near the AdS boundary, $z=0$:
\begin{equation}
\begin{aligned}
\Phi(x,z) &= \phi_1(x)z + \phi_2(x)z^2 + \cdots, \\
\eta(x,z) &= \eta_1(x)z + \eta_2(x)z^2 + \cdots, \\
A_t(x,z) &= \mu(x) - \rho_A(x)z + \cdots, \\
B_t(x,z) &= \mu_B - \rho_B(x)z + \cdots,
\end{aligned}
\label{eq:app_boundary_expansions}
\end{equation}
where $\phi_1=\eta_1=\mu_B=0$ and $\mu(x)=\mu[1+\lambda\cos(kx)]$.
After having numerically obtained the solutions by solving the equations of motion for the backgrounds dual to different phases, we further expand these quantities in Fourier modes along the $x$ direction:
\begin{equation}
\begin{aligned}
\phi_2(x) &= \phi_2^{(1/2)}\cos\!\left(\frac{kx}{2}\right)
           + \phi_2^{(3/2)}\cos\!\left(\frac{3kx}{2}\right) + \cdots, \\
\eta_2(x) &= \eta_2^{(0)} + \eta_2^{(1)}\cos(kx)
           + \eta_2^{(2)}\cos(2kx) + \cdots, \\
\rho_A(x) &= \rho_A^{(0)} + \rho_A^{(1)}\cos(kx)
           + \rho_A^{(2)}\cos(2kx) + \cdots, \\
\rho_B(x) &= \rho_B^{(1/2)}\cos\!\left(\frac{kx}{2}\right)
           + \rho_B^{(3/2)}\cos\!\left(\frac{3kx}{2}\right) + \cdots.
\end{aligned}
\label{eq:app_fourier_expansions}
\end{equation}

\begin{figure}[htp]
\centering
\includegraphics[width=0.32\textwidth]{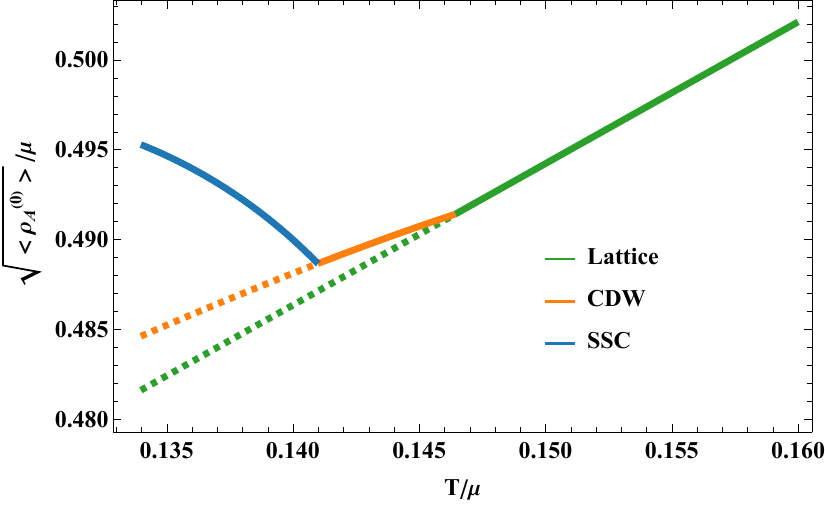}
\includegraphics[width=0.32\textwidth]{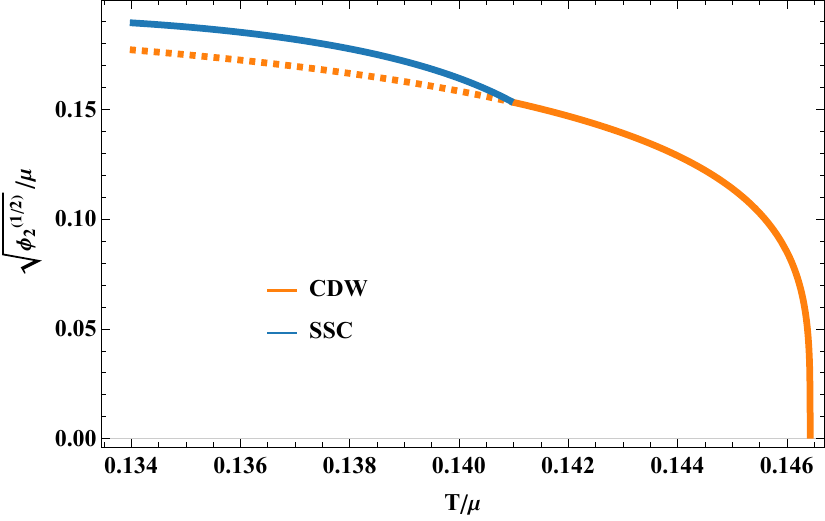}
\includegraphics[width=0.32\textwidth]{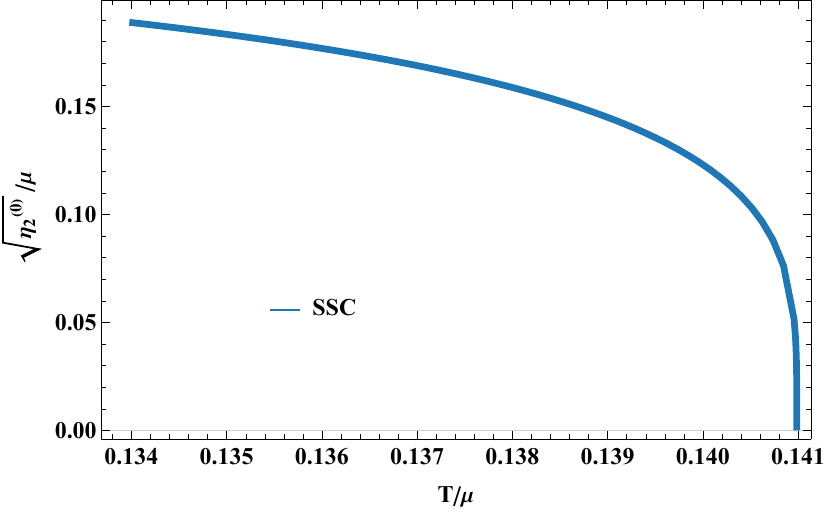}
\\
\includegraphics[width=0.32\textwidth]{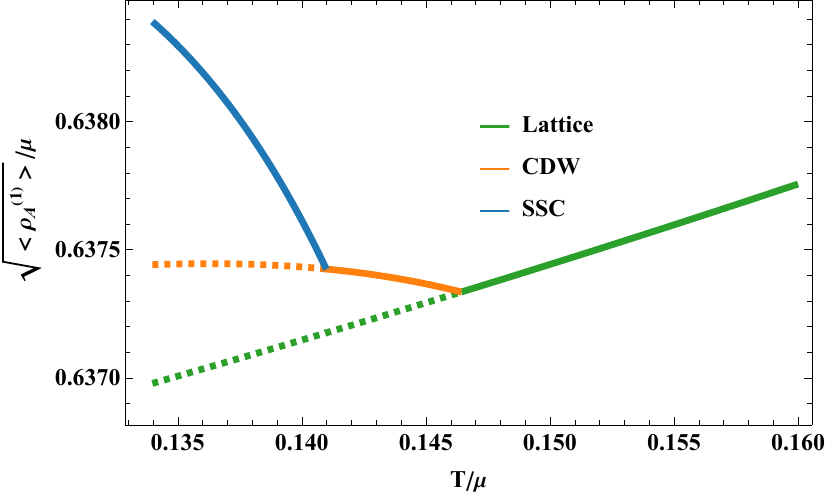}
\includegraphics[width=0.32\textwidth]{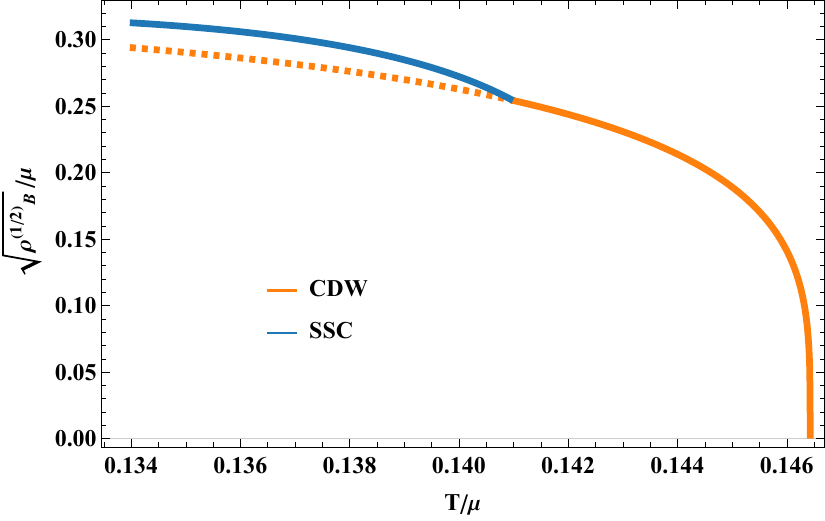}
\includegraphics[width=0.32\textwidth]{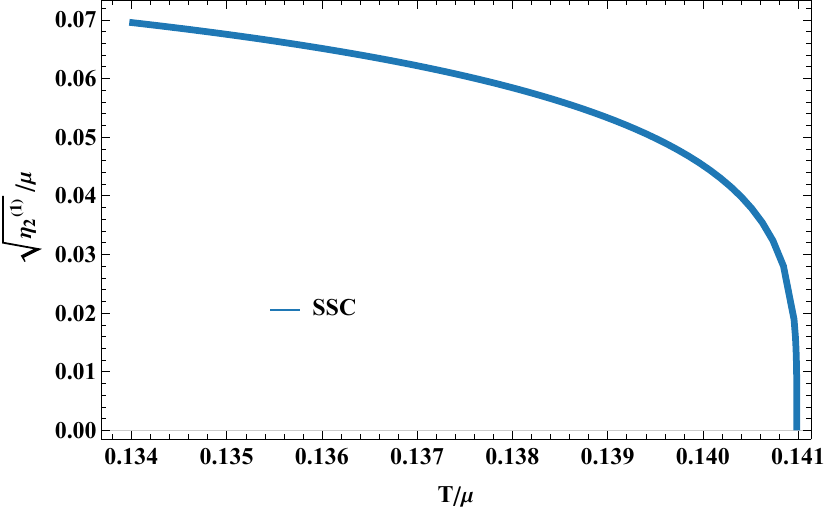}
\caption{The order parameters and the charge density as a function of temperature along the CDW-to-SSC pathway in the unlocked case with $e=3.4$ and $\lambda=0.4$. Solid lines denote the thermodynamically preferred branch in each temperature range.}
\label{fig:app_e34_l04}
\end{figure}

\begin{figure}[htp]
\centering
\includegraphics[width=0.32\textwidth]{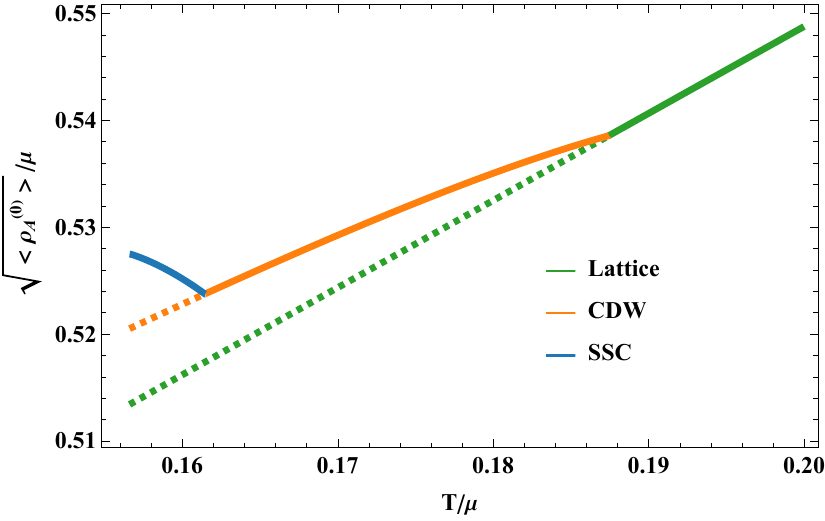}
\includegraphics[width=0.32\textwidth]{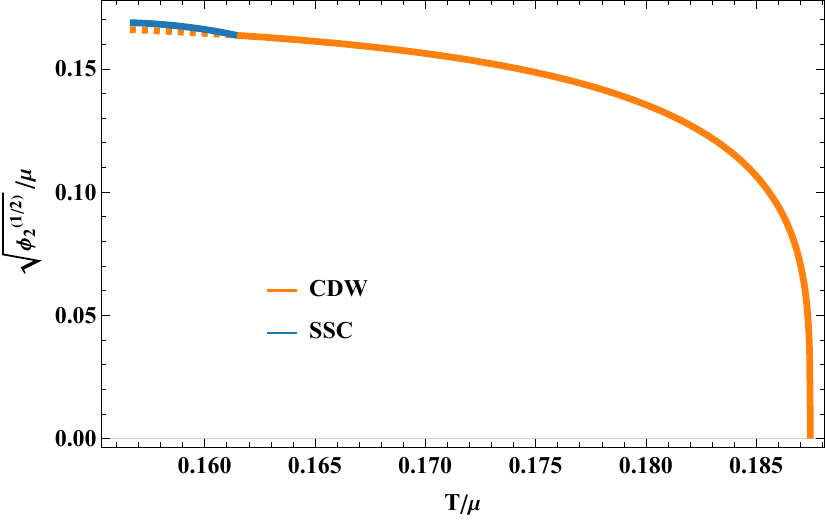}
\includegraphics[width=0.32\textwidth]{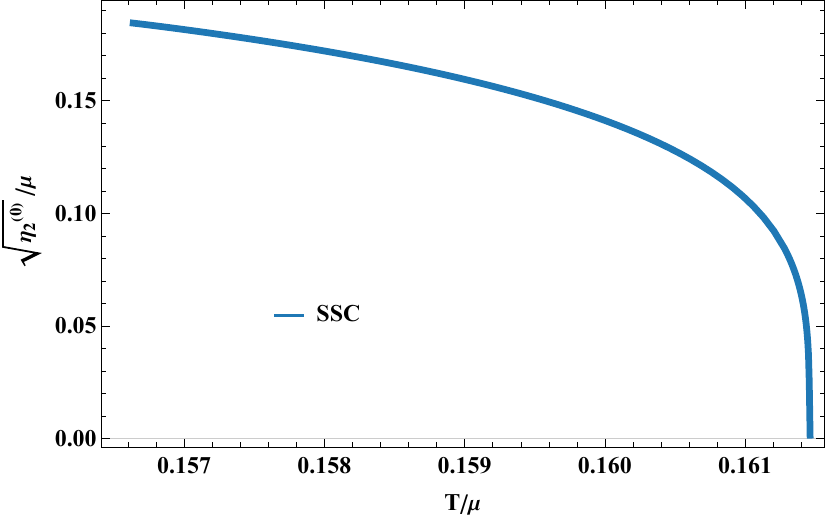}
\\
\includegraphics[width=0.32\textwidth]{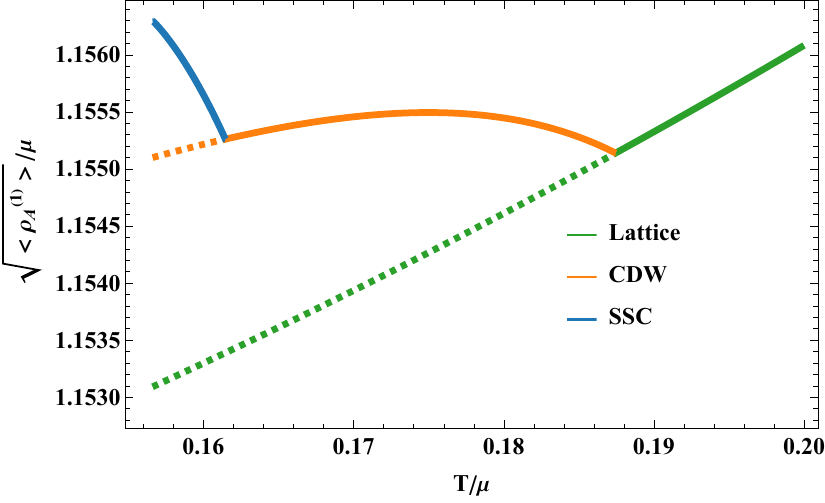}
\includegraphics[width=0.32\textwidth]{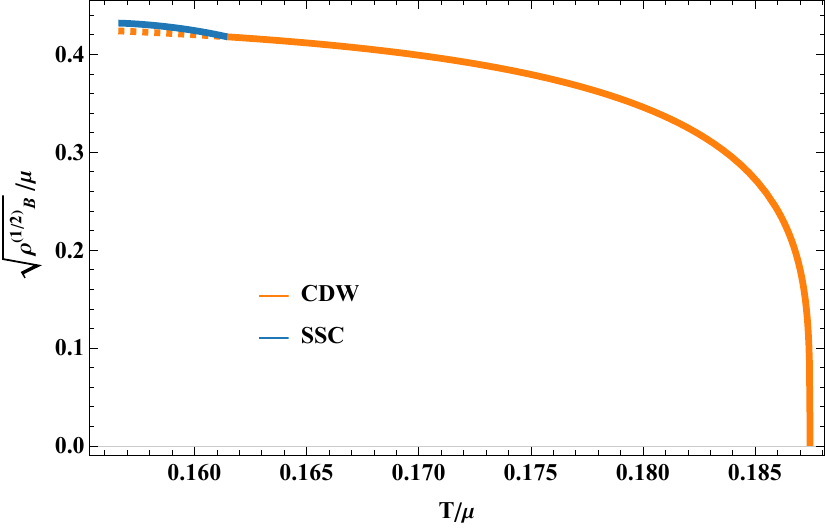}
\includegraphics[width=0.32\textwidth]{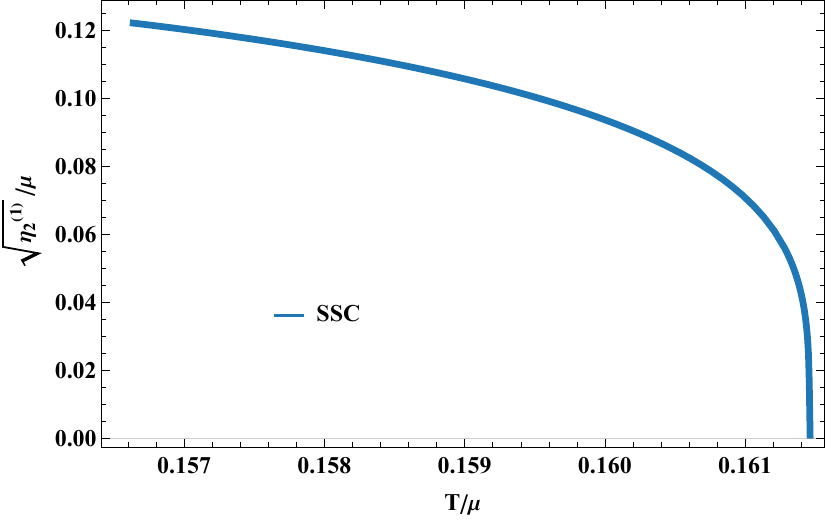}
\caption{The order parameters and the charge density as a function of temperature along the CDW-to-SSC pathway in the locked case with $e=3.4$ and $\lambda=1.3$. Solid lines denote the thermodynamically preferred branch in each temperature range.}
\label{fig:app_e34_l13}
\end{figure}

We consider two pathways to the SSC phase as studied in the main text, namely the CDW $\to$ SSC pathway with $e=3.4$ and the SC $\to$ SSC pathway with $e=3.9$. For each pathway, we show the temperature dependence of the order parameters and charge densities for both the unlocked and locked cases. Their changes along the CDW $\to$ SSC pathway are illustrated in Figs.~\ref{fig:app_e34_l04} and~\ref{fig:app_e34_l13}. It is observed that the CDW order parameter $\phi_2^{(1/2)}$ and the accompanying charge density component $\rho_B^{(1/2)}$ become nonzero below $T_{\mathrm{CDW}}$. Upon further cooling below $T_{\mathrm{SSC}}$, the uniform and modulated components of the superconducting condensate, $\eta_2^{(0)}$ and $\eta_2^{(1)}$, appear and grow. Compared with the CDW branch at the same temperature, the SSC phase has larger $\phi_2^{(1/2)}$ and $\rho_B^{(1/2)}$, indicating that superconducting condensation is accompanied by an enhancement of the existing charge modulation. At a given temperature, both $\rho_A^{(0)}$ and $\rho_A^{(1)}$ are largest on the SSC branch, intermediate on the CDW branch, and smallest on the branch with only the ionic lattice.

\begin{figure}[htp]
\centering
\includegraphics[width=0.32\textwidth]{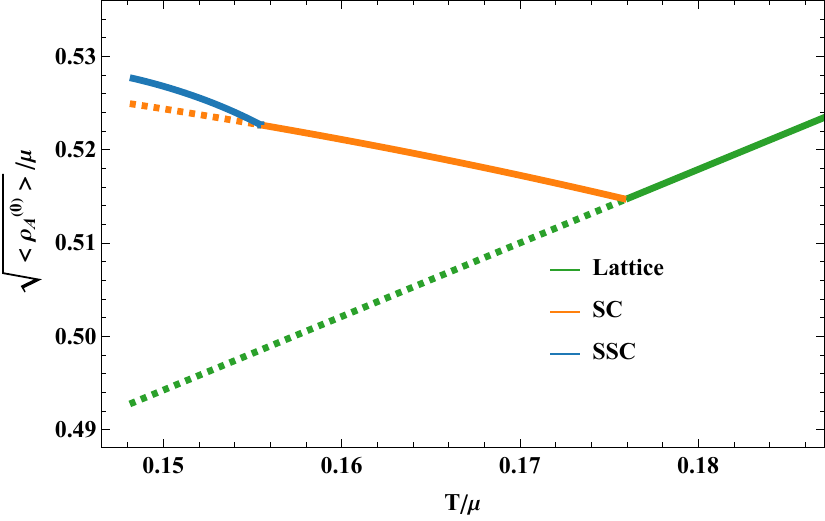}
\includegraphics[width=0.32\textwidth]{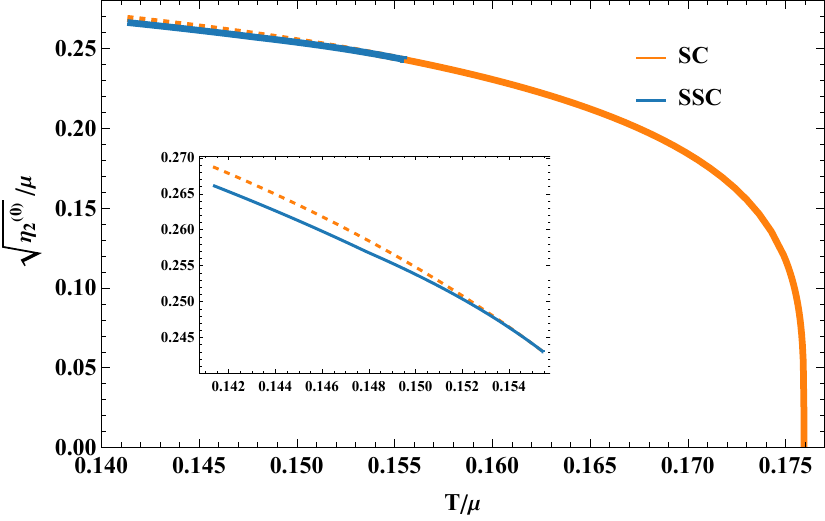}
\includegraphics[width=0.32\textwidth]{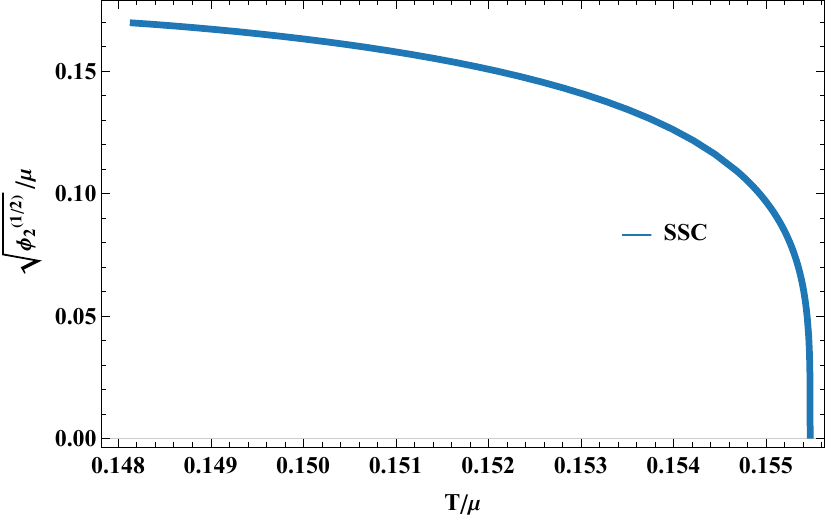}
\\
\includegraphics[width=0.32\textwidth]{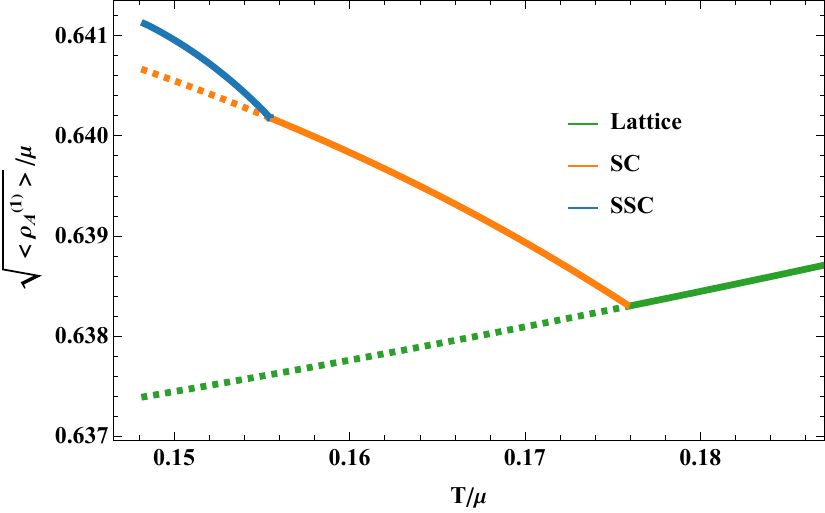}
\includegraphics[width=0.32\textwidth]{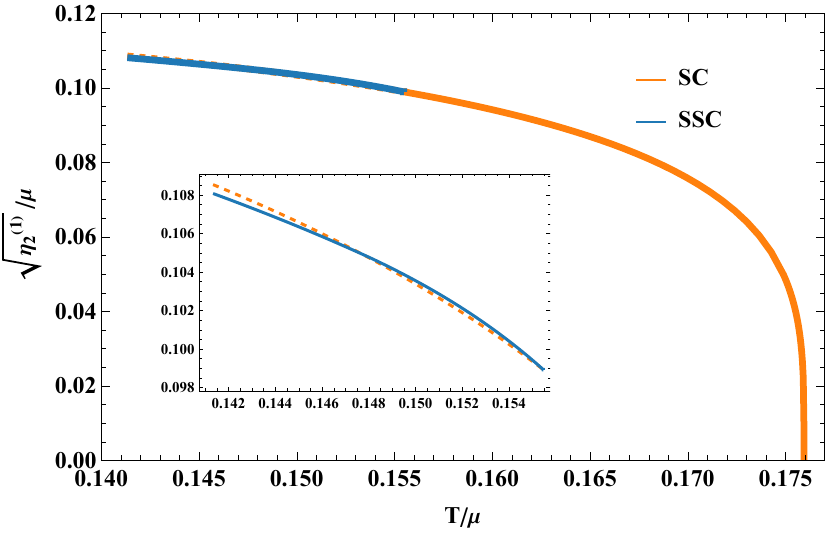}
\includegraphics[width=0.32\textwidth]{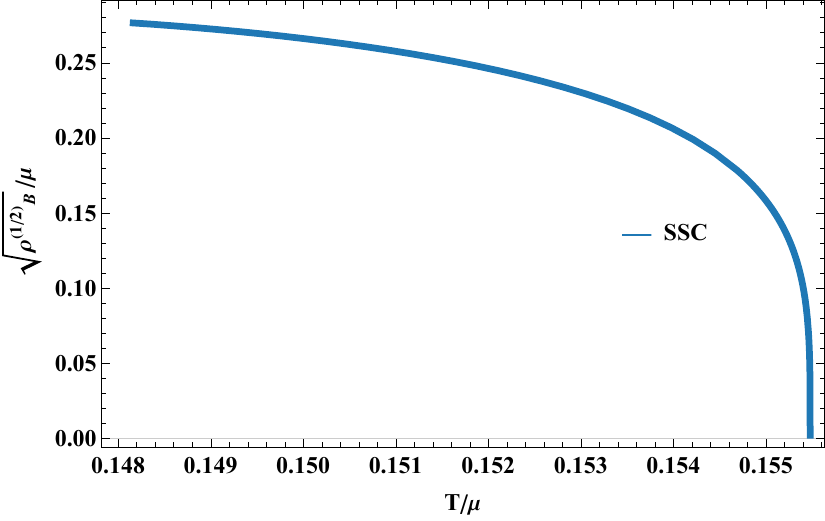}
\caption{The order parameters and the charge density as a function of temperature along the SC-to-SSC pathway in the unlocked case with $e=3.9$ and $\lambda=0.4$. Solid lines denote the thermodynamically preferred branch in each temperature range.}
\label{fig:app_e39_l04}
\end{figure}

\begin{figure}[htp]
\centering
\includegraphics[width=0.32\textwidth]{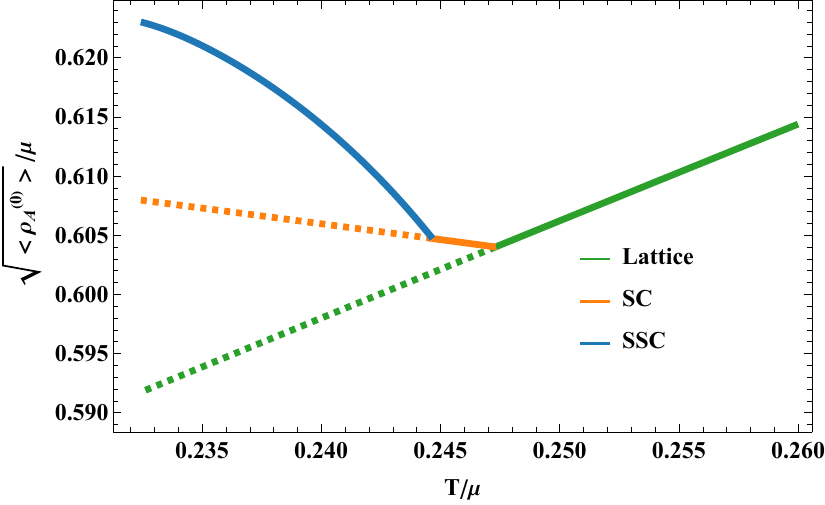}
\includegraphics[width=0.32\textwidth]{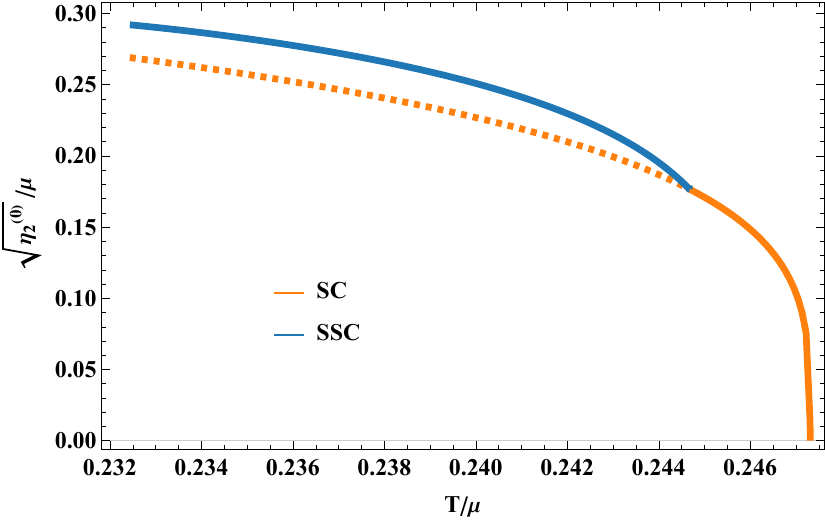}
\includegraphics[width=0.32\textwidth]{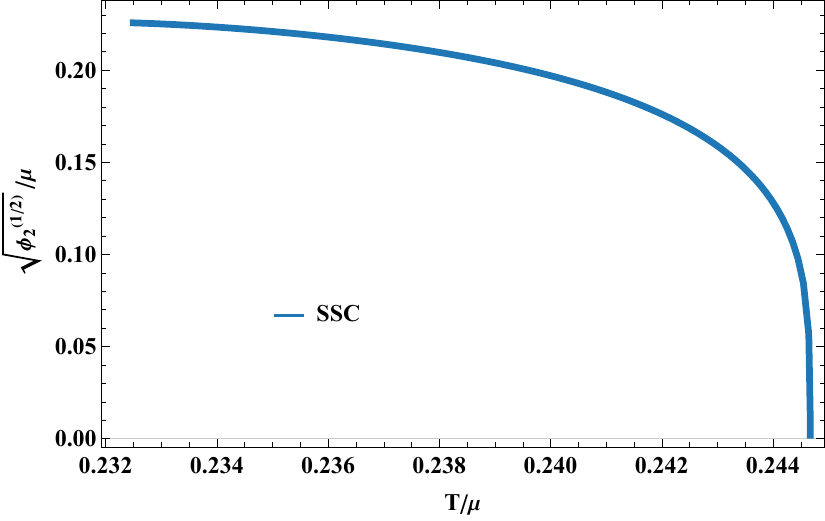}
\\
\includegraphics[width=0.32\textwidth]{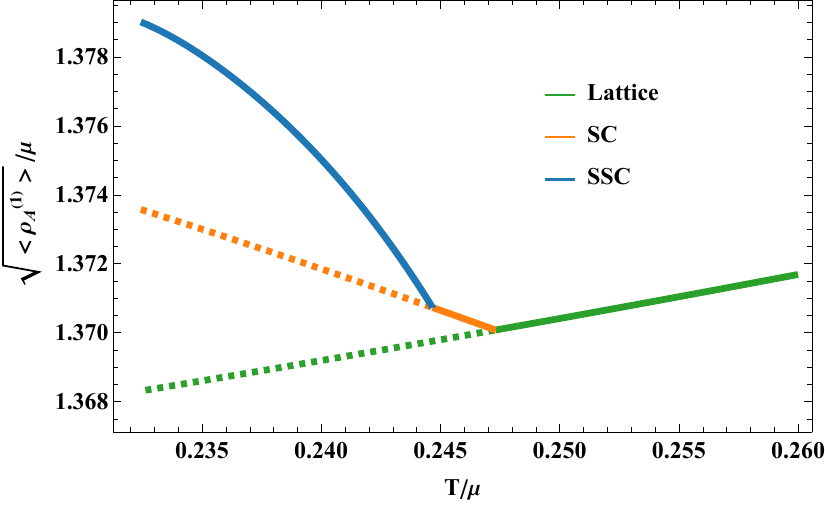}
\includegraphics[width=0.32\textwidth]{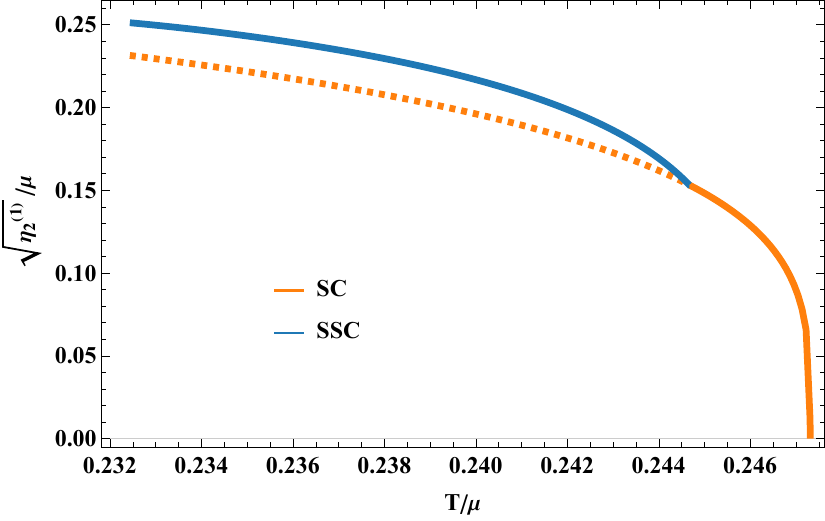}
\includegraphics[width=0.32\textwidth]{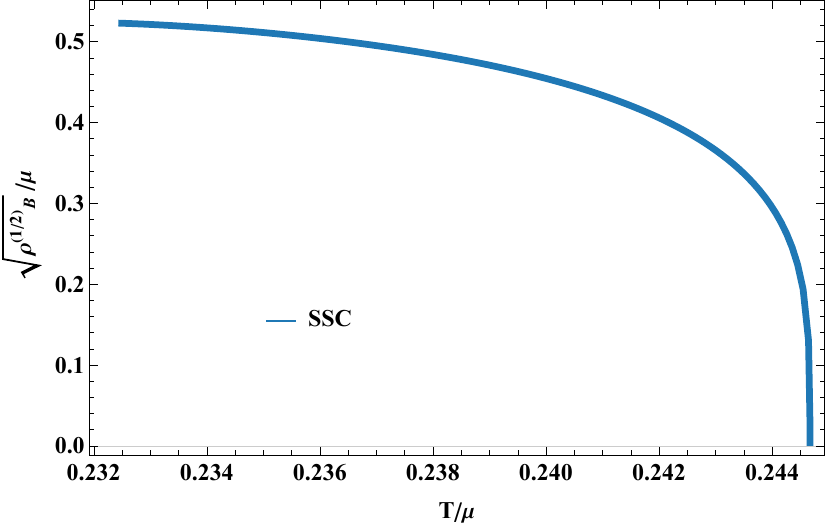}
\caption{The order parameters and the charge density as a function of temperature along the SC-to-SSC pathway in the locked case with $e=3.9$ and $\lambda=1.8$. Solid lines denote the thermodynamically preferred branch in each temperature range.}
\label{fig:app_e39_l18}
\end{figure}

Their changes along the SC $\to$ SSC pathway are illustrated in Figs.~\ref{fig:app_e39_l04} and~\ref{fig:app_e39_l18}. One observes that $\eta_2^{(0)}$ and $\eta_2^{(1)}$ become nonzero below $T_{\mathrm{SC}}$, whereas $\phi_2^{(1/2)}$ and $\rho_B^{(1/2)}$ appear only upon further cooling below $T_{\mathrm{SSC}}$. In the unlocked case ($\lambda=0.4$), the SSC and SC values of $\eta_2^{(0)}$ and $\eta_2^{(1)}$ are close, with $\eta_2^{(0)}$ slightly smaller in the SSC phase. In the locked case ($\lambda=1.8$), both components are larger in the SSC phase than in the SC phase at the same temperature. In addition, $\eta_2^{(0)}$ is much larger than $\eta_2^{(1)}$ in the unlocked case, whereas these two quantities are comparable in the locked case, with $\eta_2^{(1)}$ remaining smaller. For the charge density components $\rho_A^{(0)}$ and $\rho_A^{(1)}$, the values at a given temperature are largest on the SSC branch, intermediate on the SC branch, and smallest on the branch with only the ionic lattice.

\clearpage
\bibliographystyle{style1}
\bibliography{scalar}

\end{document}